# Electron-Induced Radiation Chemistry in Environmental Transmission Electron Microscopy

*Kunmo Koo*[1,2], *Nikhil S. Chellam*[3,4], *Sangyoon Shim*[1], *Chad A. Mirkin*[1,3,4], *George C. Schatz*[3,4], *Xiaobing Hu*[1,2]*, and *Vinayak P. Dravid*[1,2,4]*

[1]Department of Materials Science and Engineering, Northwestern University, Evanston, IL 60208, United States

[2]The NU*ANCE* Center, Northwestern University, Evanston, IL 60208, United States

[3]Department of Chemical & Biological Engineering, Northwestern University, Evanston, IL 60208, United States

[4]International Institute for Nanotechnology, Northwestern University, Evanston, IL 60208, United States

ABSTRACT

Environmental transmission electron microscopy (E-TEM) enables direct observation of nanoscale chemical processes crucial for catalysis and materials design. However, the high-energy electron probe can dramatically alter reaction pathways through radiolysis – the dissociation of molecules under electron beam irradiation. While extensively studied in liquid-cell TEM, the impact of radiolysis in gas-phase reactions remains unexplored. Here, we present a numerical model elucidating radiation chemistry in both gas and liquid E-TEM environments. Our findings reveal that while gas-phase E-TEM generates radiolytic species with lower reactivity than liquid-phase systems, these species can accumulate to reaction-altering concentrations, particularly at elevated pressures. We validate our model through two case studies: the radiation-promoted oxidation of aluminum nanocubes and disproportionation of carbon monoxide. In both cases, increasing the electron beam dose rate directly accelerates their reaction kinetics, as demonstrated by enhanced $AlO_x$ growth and carbon deposition. Based on these insights, we establish practical guidelines for controlling radiolysis in closed-cell nanoreactors. This work not only resolves a fundamental challenge in electron microscopy but also advances our ability to rationally design materials with sub-Ångström resolution.

## Introduction

The ability to directly observe and control chemical reactions at the nanoscale is crucial for advancing fields from catalysis to energy storage. Environmental transmission electron microscopy (E-TEM) has emerged as a powerful tool for this purpose, offering sub-Ångström resolution of dynamic chemical processes. However, the high-energy electron beam (often assumed an agnostic probe) can simultaneously interact with the specimen of interest.[1] Although electrons in TEM (when accelerated to 30-300 keV) have lower energies than those of conventional broad-beam radiation sources (e.g., X-rays, γ-rays, and neutrons), their concentrated flux results in greater localized energy transfer. Subsequent electron excitation and ejection from both the specimen and its surroundings result in two predominant damage mechanisms: physical atom ejection (knock-on damage) and the generation of reactive radicals and ionized molecules that initiate reaction cascades (radiolysis).[2] As the ionization cross-section for molecules increases at lower acceleration energies,[3] radiolysis effects are critical to consider amongst the vast array of electron-fluid interaction mechanisms, including knock-on damage, heating,[4, 5] and electrical charging.[1, 6]

Studies involving radiation-sensitive specimens (e.g. liquids, gases, polymers, and biomaterials) have developed methods to control radiolysis – either through the use of cryogenic temperatures to suppress radical diffusion[16-18] or scavengers to neutralize reactive species.[19-21] On the other hand, the ostensibly destructive nature of electron beam radiolysis has, perhaps counterintuitively, proven valuable for facilitating chemical reactivity in certain instances. To this end, researchers have successfully exploited radiolysis to interrogate battery material transformations[7-9] as well as the growth[10-12] and etching of nanocrystals[13-15] with atomic-scale detail. These advances underscore how the precise control of the radiation environment is necessary for any quantitative analyses of chemical reactions in E-TEM.

Despite their importance in many chemical processes, investigations of gas-phase systems (GPTEM) remain comparatively limited.[2, 22, 23] Our current understanding of radiation chemistry in E-TEM is primarily based on Schneider *et al.*'s liquid-phase (LPTEM) model which describes the temporal evolution of chemical species under homogeneous electron beam exposure within aqueous environments.[2] This foundational work has been progressively refined to account for various physicochemical phenomena including the addition of reaction systems,[21, 24-26] pressure-limited solubility,[27] and kinetic sensitivity factors[28] arising from backscattered and secondary electrons (BSEs), alongside the integration of temperature[29] and flow rate dependencies[30] to enhance predictive accuracy. However, the direct application of liquid-phase radiolysis models to gas-phase systems fundamentally misrepresents their underlying physics. The Damköhler number (Da) – i.e., the relative rates of reaction to diffusion – varies by several orders of magnitude between liquids (Da ~ $10^1$ to $10^6$) and gases (at 10 torr, Da ~ $10^{-10}$ to $10^{-6}$). Consequently, LPTEM assumes homogeneous radiation chemistry within the illumination window and often neglects diffusion effects altogether – an assumption invalid for GPTEM.[25] This can be partially overcome *via* extant finite element tools which, nonetheless, cannot thoroughly simulate electron beam effects such as those from a scanning TEM (STEM) probe.

Herein, we present a generalized numerical model for analyzing radiation chemistry within closed cell-type E-TEM **(Figure 1A)**. This implementation accommodates both S/TEM imaging conditions in isobaric, room temperature (25 ºC) liquid and gaseous media, with the ability to account for diffusion dynamics. From this model, we determine the steady-state concentrations of radiolytic species and estimate the propensity of redox reactions with electron dose rate. Our investigation reveals significant distinctions in radiolytic species accumulation between atmospheric GPTEM and low-pressure open-cell E-TEM.[31, 32] These findings suggest that the use of a scanning electron probe more efficiently mitigates gas-phase radiolysis (due to their high diffusivities) compared to a stationary TEM beam. To validate our model, we conducted real-time experiments tracking electron dose-accelerated processes: namely, the oxidation of aluminum nanoparticles **(Figure 1B, S1)** and disproportionation of carbon monoxide, where increasing the electron beam dose rate concomitantly increases both $AlO_x$ layer growth and the carbon deposition rate.

## Results and Discussion

### Electron Radiolysis Model for Gas- and Liquid Phase TEM

Modern high-end probe corrected microscopes demonstrate remarkable energy deposition capabilities. A typical 200 keV instrument (with a 1 Å diameter probe and a 62 pA current) can dose specimens with rates approaching $1.95 \times 10^{15}$ Gy/s under a stationary probe position. Even under scanning conditions at undersampled magnifications, dose rates remain impressively high (ca. $10^7$-$10^9$ Gy/s) – substantially exceeding radiation levels observed in X-rays[38, 39] (few Gy/min), $^{60}$Co γ-irradiation[40] (Gy/s) and high-flux synchrotron X-rays[41-43] (kGy/s).

In closed-cell E-TEM, specially designed holders enclose the specimen between two electron-transparent windows (e.g. C or $SiN_x$) and confine the fluid to the sample.[33, 34] This technique is categorized by the electron irradiation mode and the phase of the medium: LP(S)TEM and GP(S)TEM. The electron dose mode affects the types of species initially generated and their spatial distribution. TEM mode typically employs static electron irradiation, utilizing higher currents (several nA) over large areas ($10^1$-$10^4$ nm). Conversely, scanning TEM (STEM) mode operates with significantly reduced probe currents (several pA) focused over smaller areas (< 1 Å). Electrons are distributed across the region of interest (ROI) after brief dwell times of several microseconds. Consequently, radiation chemistry in E-TEM varies by illumination mode, as each mode generates distinct reaction dynamics and steady-state species concentrations depending on beam rasterization and sampling methods.

High-energy electron interactions with atoms produce low-energy BSEs, whose interaction mechanisms vary dramatically between liquid and gas phases. In liquid environments, the spur effect describes how BSEs, unable to travel far from the irradiation source, generate dense ion pair clusters along their trajectories, thereby creating localized increases in the concentration of radiolysis species. In the gas phase, however, the BSE's inelastic mean free path (IMFP) is far greater than the observation area, effectively precluding any spurring.[35] For example, for 100 eV electrons, the inelastic cross section ($\sigma_i$) for a low-Z element is $10^{-20}$ $m^2$,[36] while the density of atoms ($n$) under 10 torr pressure is on the order of $10^{23}$ $m^{-3}$, yielding an IMFP ($\lambda_i$) of $10^{-3}$ m ($\lambda_i = 1/n\sigma_i$). Consequently, gas-phase dose rate calculations can more readily adapt broad-beam radiation chemistry kinetic models with fewer compensatory assumptions.

There are two conventions to describe the electron dose rate $\varphi$, each providing different aspects of electron-sample interactions. The unit $e^-/Å^2 \cdot s$ quantifies the number of electrons impinging on a unit area per second, which affects the optical performance and signal-to-noise ratio in microscopy. This can be mathematically represented as

$$\varphi \left[\mathrm{e^-/Å^2 s}\right] = \frac{I}{q_e A} \tag{1}$$

where $I$ [A] denotes the beam current, $A$ [Å$^2$] is the illumination area, and $q_e$ $[C]$ is the fundamental electron charge. Alternatively, the unit Gy/s quantifies the energy absorbed per unit mass (J/kg) per second, which conveys how radiation affects the sample, and is described by

$$\varphi \text{ [Gy/s]} = \frac{10^5 SI}{A} \tag{2}$$

where $S$ [MeV cm$^2$/g] is the material's intrinsic electron stopping power **(Supplementary Text I)** and $A$ [m$^2$] is the illumination area.[2] STEM dose calculations introduce additional complexity. To calculate the electron dose in STEM mode ($\varphi_S$), factors such as dwell time, sampling size, and pixel counts must be considered in Equations (1) and (2) as they determine the degree of oversampling. In this work, we used a 1:10 undersampling ratio for our calculations and experiments, requiring a reduction in dose rate by a factor of 100 ($\varphi_S = \varphi t_d / 100 t_f$) at the irradiated position,[37] where $t_d$ and $t_f$ denote the dwell and frame times, respectively. The subsampling ratio used herein represents an intermediate STEM magnification for a typical *in situ* experiment using a 512 × 512 nm$^2$ viewing area with a 512 × 512 pixel resolution (**Figures 1B-D**).

The concentration of molecular species generated through radiolysis intimately correlates with the surrounding fluid's molecular density, which, in turn, is contingent on phase. The initial concentration of species generated by the electron beam ($R_i$) can thus be estimated as

$$R_i = \frac{\rho_B \varphi G}{N_A} \text{ [M/s]} \tag{3}$$

where $\rho_B$[kg/L] is the density of the surrounding medium $B$, $\varphi$ [Gy/s] is the electron dose rate, $G$ [species/$J$] is the statistical number of generated species following energy absorption per Joule

**(Supplementary Text I)**, and $N_A$ is Avogadro's constant.[2] Critically, the diffusivity of species $i$ in medium $j$ ($D_{ij}$) also depends on the fluid density; that is,

$$D_{ij} = \frac{0.00143 \times T^{1.75}}{\sqrt{M_{ij}}\left(V_i^{1/3} + V_j^{1/3}\right)^2 p} \tag{4}$$

where $T$ [K] is temperature, $p$ [atm] is pressure, $M_{ij} = 2/(M_i^{-1} + M_j^{-1})$ is the effective species mass, and $V$ is the molecular diffusion volume **(Supplementary Text II)**.[44-46] For example, from Equation (4), the diffusivity of the hydroxyl radical (OH·) in 10 torr water vapor is approximately $1.1 \times 10^{15}$ nm$^2$/s. However, the diffusivity of OH· in liquid water is only $2.8 \times 10^9$ nm$^2$/s – a difference of six orders of magnitude.[47] We incorporate Equations (3) and (4) within a modified version of the constitutive reaction-diffusion equation to evaluate the spatiotemporal concentrations of species $i$, where[2]

$$\frac{\partial C_i}{\partial t} = R_i - \sum_j k_{ij} C_i C_j + \sum_{j,k \neq i} k_{jk} C_j C_k + \frac{\partial}{\partial x}\left(D_i \frac{\partial C_i}{\partial x} - \mathrm{v} C_i\right) \tag{5}$$

which couples the initial generation of species *i* with mass transport. Additionally, the second and third terms on the right-hand side (RHS) represent the consumption and production of *i via* chemical reactions with rate constants $k_{ij}$ and $k_{jk}$, respectively **(Supplementary Text III)**. The fourth term on the RHS represents mass transport due to advection and diffusion.

We have implemented this coupled partial differential equation (PDE) in MATLAB[48] (**Supplementary Text IV-V**) to assess the diffusion-dominated ($\mathrm{v} = 0$) electron dose-induced reaction profile within these closed-cell nanoreactors **(Figure 2)**.[49] We compare our results with those obtained using the model proposed by Schneider, *et al.*[2] under identical dose rates and fluid conditions with a low diffusivity ($D_i$ = 1 $nm^2$/s; **Figure S2**). Using our model, the steady-state concentrations (assessed at $10^{-2}$ s) deviate by ~2.5% for higher concentration species (e.g., $H_2$, $H_2O_2$, and $O_2$) and ~10% for lower concentration species (e.g., $e^-$, OH· and H·). This slight deviation is related to the numerical solver (pdepe) we used to solve the PDE. In this method, assigning $D$ to 0 would cause the concentration and source terms (LHS and 2nd term of RHS in Equation S5) to diverge. Consequently, to benchmark against previous models that do not account for diffusion, we evaluate Equation (5) in the limit of small $D_i$ ~1 $nm^2$/s. By doing so, the coefficients ($1/D_i$) in the differential and source terms in Equation S5 significantly increase, thereby allowing for an increased error tolerance (Equation S6 and S8). Under realistic experimental conditions, the spatial profiles for $H_2$, $O_2$, $e^-$, and $H^+$ agree well with previously-reported heterogeneous cases calculated using a finite element solver within a timestep of $10^{-4}$ to $10^{-2}$ s **(Figure S3)**.[2]

The spatiotemporal distributions of each radiolytic species in water under typical high-magnification TEM conditions (with a probe current of 4.2 nA, diameter of ~400 nm, and calculated dose rate of ~$9.4 \times 10^9$ Gy/s) are shown in **Figure 2A-B** and **Movie S1**. Following electron irradiation, each species propagates beyond the ROI and equilibrates after $10^{-1}$ s. The affected area extends by several micrometers for higher concentration species ($H_2$, $H_2O_2$, and $O_2$), while lower concentration species (OH·, $e^-$, H·) diffuse only by a few hundred nanometers. The concentrations in GPTEM are shown in **Figure 2C** and **Movie S2**. Species in LP- and GP-TEM significantly differ in their steady-state concentration due to differences in their molecular density. In particular, the steady-state concentrations of OH·, H·, and $H_2$ are reduced by factors of ~$10^6$ to $10^{10}$ in liquid water compared to 10 torr water vapor due to decreased diffusivities and reaction kinetics.

Electron dose-promoted reactions vary in STEM mode due to continuous electron probe scanning with distinct effects in liquids and gases **(Figure 2D, E)**. Under a single electron probe (62 pA current, 1 Å diameter; ~$6.7 \times 10^9$ Gy/s dose rate at 3 μs dwell time), the species concentrations equilibrate within one microsecond, reaching concentration levels up to ~10 to $10^3$ times higher than those in TEM mode within the illuminated area **(Movie S3)**. After their concentrations stabilize, the probe then moves to adjacent pixels, causing a stepwise decrease in their radial concentration profile **(Movie S4)**. The initial radiolysis species accumulate in liquids but disperse readily in gases due to their different diffusivities. This, then, leads to GPSTEM steady-state concentrations to be approximately equivalent to those in GPTEM. For instance, the steady-state concentration of hydroxyl radicals is the highest in LPSTEM ($10^4$-$10^5$ μM) and LPTEM ($10^3$ μM), then GPSTEM ($10^{-4}$ μM) and GPTEM ($10^{-4}$ μM).

Concentration fluctuations under different illumination modes are less pronounced for $H_2O_2$, $H_2$, and $O_2$ compared to OH·, H·, and $e^-$. The first group either has a low *G*-value or is reaction-generated; direct electron dosage minimally affects their concentrations compared to the latter. In GP(S)TEM, initial generation rates determine the steady-state radiolytic species, as evidenced by the Damköhler numbers for initial generation ($10^2$-$10^4$) *vs.* reaction kinetics ($10^{-10}$-$10^{-6}$) at 10 torr (Equation S17, **Supplementary Text V**). In gaseous conditions, however, species generated by direct electron irradiation (OH·, H·, O·, $H_2$) accumulate within the illumination area, showing a strong *G*-value dependence per Equation (3), with reaction kinetics slightly influencing their steady state concentrations per Equation (5). These radiolytic species are evenly distributed throughout the imaging area **(Figures S4-S5)** due to their high gas-phase diffusivities, affecting regions beyond the imaging area. While reactive species concentrate in the viewing area in LPSTEM, species in GPSTEM rapidly diffuse from the point source, resulting in lower concentrations than TEM in remote areas or non-illuminated pixels **(Figure 2C, 2F)**.

In conventional reactors, chemical species continuously flow in and out through mass transfer. Recent studies indicate that higher liquid flow rates can thus reduce the production of radiolytic species *via* washout.[30] Thus, we evaluated any 1-D advection effects ($v \neq 0$) in both liquid and gaseous media **(Supplementary Text V)**. Typical flow rates in LPTEM and GPTEM are several μL/min and 1 sccm, which yield average linear velocities of approximately 0.01 m/s and 1 m/s, respectively. However, the actual linear velocities in the observation area are often slower due to bypass fluid paths and flow resistance within the narrow channels. Our analysis using Equation (5) suggests that while advection may influence liquid-phase dynamics **(Figures S6-S7)**, gas radiolysis is minimally impacted (up to 1200 sccm flow rates; **Figure S8**).

**Electron Radiolysis Promoted Aluminum Oxidation**

Al nanocubes are an excellent test case to validate our model. Their {100} facets and large sizes (200 nm) enable facile measurement as they stand upright on the membrane and can be oxidized from $Al^0$ to $Al^{3+}$ without additional stimuli (e.g. heat) and significant facet dependent kinetics. Similar to bulk Al,[50] these nanocubes have a ~3 nm native oxide layer that inhibits bulk oxidation under ambient conditions **(Figure S9)**. However, electron beam irradiation dramatically accelerates oxide growth in oxygen-rich environments ($O_2$, $CO_2$, air, and $H_2O$; **Figures 1B** and **S10**). These findings align with a prior report showing accelerated reactions under $0.94\text{-}2.85 \times 10^{-3}$ Gy of γ-ray radiation in atmospheric air.[51] The electron-induced oxidation rate (1 Å/s) far exceeds that in radiation-free conditions (3 nm over several days)[50, 52] and increases proportionally with electron dose rate **(Figure 1B, Movie S5-7)**. Notably, the oxide layer does not grow in the absence of oxygen-containing species (*i.e.,* vacuum, $N_2$, or Ar), highlighting electron dose effects in high-pressure gaseous redox reactions, contrary to what is observed in low pressure open-cell GPTEM.

Al oxidation products vary by environment.[31] The oxidation products in $H_2O$ vapor yields a mixture of aluminum (III) oxide ($Al_2O_3$), aluminum hydroxide ($Al(OH)_3$), and oxyhydroxides,[53] while $O_2$, $CO_2$, and air primarily yield $Al_2O_3$. Indeed, we find the oxidized product morphology to be quite different under water vapor, showing distinct, porous morphologies **(Figure S11)**. Though precise product identification proves challenging using TEM-based methods (e.g., ED, EDS, EELS) due to their excessive thickness **(Figure S12**), the pronounced structural differences observed herein lead us to avoid a direct comparison of Al redox kinetics with water vapor and other gases ($O_2$, $CO_2$, and air).

Under our tested STEM conditions, oxidation proceeds at several atomic layers per second (**Figure 3**) following pseudo-zeroth order kinetics. These results suggest that reaction site availability is a rate-determining factor, while electron radiolysis continuously replenishes any consumed species. This rapid growth produces poorly crystalline $AlO_x$ shells that minimally impede species diffusion through the shell to the reaction front. Consequently, these oxide shells can easily reach 10-70 nm thicknesses and are 2–3 times larger than under extreme conditions without radiation (**Figure 4**). At a fixed dose rate of 75.53 $e^-/Å^2 \cdot s$ (~$3.4 \times 10^8$ Gy/s), the oxidation rates are the fastest for $O_2$ (1.12 Å/s), followed by dry air (0.62 Å/s) and slowest for $CO_2$ (0.38 Å/s). Moreover, the oxidation rate linearly increases with electron dose, with the steepest slope for pure $O_2$.

Electron radiolysis within these high-pressure gaseous environments were calculated using several oxygen-containing gases: i.e., $O_2$, $CO_2$, and dry air (78.5% $N_2$ + 21.5% $O_2$). The primary radiolysis yields were determined as:[54-57]

$$O_2 \rightarrow e^-, O^+, O_2^+, O\cdot \quad (1)$$

$$CO_2 \rightarrow e^-, O\cdot, O^+, CO, CO^+, CO_2^+ \quad (2)$$

$$Air \rightarrow e^-, O^+, O_2^+, O\cdot, N_2^+, N^+, N\cdot \quad (3)$$

These initial species trigger subsequent reaction cascades (discussed in **Supplementary Text III**). In GPSTEM, radiolytic species concentrations monotonically rise and quickly approach steady state. Similar to our observations under 10 torr water vapor (*vide supra;* **Figure 2F**), only initially generated species accumulate within the viewing area when $O_2$, $CO_2$, and air are introduced (**Figure 5**).

Among the accumulated species from radiolysis pathways (1) to (3), both $O_2^+$ and $O^+$ serve as active oxidizers for the conversion of $Al^0$ to $Al^{3+}$.[58] Oxygen radicals ($O\cdot$) have a high oxidizing capability ($E^0$ = 2.42 V), while free electrons ($e^-$; $E^0$ = -2.9 V) are the only reductive species that can effectively accumulate.[59] Consequently, the ratio between the oxidative species present and the reductive free electron concentration ($\Xi$) can provide insight into the oxidation rates across different gases:

$$\Xi = \frac{[O_2^+] + [O^+] + [O\cdot]}{[e^-]} \quad (6)$$

Among these representative oxygen-containing gases, pure $O_2$ has the highest $\Xi$ of 1164.08, while dry air and $CO_2$ have a $\Xi$ of 237.5, and 50.24, respectively. This trend aligns well with our measurements, where we find Al oxidation to be the fastest in $O_2$, followed by air and slowest in $CO_2$.

In addition, the gas pressure and electron dose rate strongly affect the steady-state reactive species concentration **(Figure 6)**. The concentration of each species follows a quadratic relationship with base pressure (at a fixed electron dose) and a linear relationship with electron dose rate (at fixed pressures within 10-760 torr and currents of 30-1000 pA), described by:

$$C_i \propto \alpha p^2 \propto \beta I \tag{7}$$

Here, $\alpha$ [μM/torr$^2$] and $\beta$ [μM/pA] are fitted empirical parameters (provided in **Table 1**) that determine the concentration of species $i$ produced per unit pressure and current, respectively. The quadratic pressure dependence of $C_i$ suggests that bimolecular collision events form gaseous reactive species under electron beam irradiation, while the linear relationship with dose rate suggests that the beam current acts individually for each molecule when the fraction of ionized species is sufficiently low and double ionization events are negligible. In open-cell E-TEM, ($10^{-7}$ to $10^{-5}$ torr), radiolysis effects are often negligible as they are $10^{14}$ times lower than in closed-cell setups. Together, this implies that pressure reduction more effectively suppresses radiolysis than dose rate control in closed-cell E-TEM.

We further evaluate the dose-dependent oxidation tendency between the different species by combining Equations (6) and (7):

$$\Xi = \frac{[\mathrm{O}_2^+] + [\mathrm{O}^+] + [\mathrm{O}]}{[\mathrm{e}^-]} = \frac{\sum \beta_{\mathrm{ox}}}{\beta_{e-}} I \tag{8}$$

At 10 torr, the current coefficient ratio ($\sum \beta_{\text{ox}}/\beta_{e-}$) is the highest for $O_2$ ($1.16 \times 10^3$), followed by air ($2.37 \times 10^2$), and $CO_2$ ($5.02 \times 10^1$), in turn predicting a stronger dose dependency for oxidation by $O_2$. These relations agree with the experimentally-observed linear increase in oxidation rate and their slope with respect to electron dose.

Although these trends qualitatively agree with our physical models, we note that there remains a quantitative gap between these simulations and our observed experimental results. Growing an $Al_2O_3$ shell at a rate of 0.1 Å/s on a 200 nm Al nanocube requires approximately $1.42 \times 10^5$ O atoms to be consumed per second from the ROI (in STEM mode, 0.5 µm × 0.5 µm × 10 µm at 10 torr, yielding a gas volume of $\sim 10^{-15}$ L; **Figure S13-S14)**. However, our model predicts calculated steady state species concentrations ($O\cdot$, $O_2^+$, and $O^+$) of only ~0.1-1 molecule per ROI at 10 torr, reaching $\sim 3.8 \times 10^3$ at 1 atm. This disparity likely stems from surface adsorption effects, which reduce species diffusion as radicals accumulate at the surface, thereby providing a quantitative underestimation in our simulations. Incorporating adsorption effects *via* sensitivity factors could better align numerical predictions with our experimental results, though our model nonetheless provides valuable qualitative insights into radiolysis dynamics and other observed trends.

**Radiolytic Disproportionation of Carbon Monoxide**

To demonstrate our model's broader applicability, we examine another radiolysis case: the disproportionation reaction of CO within closed-cell GPTEM. Akin to the Boudouard reaction under nonthermal conditions, this process yields $CO_2$ and solid carbonaceous species, described by[60]

$$\mathrm{CO} \rightarrow \mathrm{CO_2}, (\mathrm{C_3O_2})_x, \mathrm{C} \tag{4}$$

Upon the introduction of CO and subsequent electron irradiation, we observe the deposition of carbonaceous materials within the viewing area **(Figure 7A)**. This observation, which was absent when other gases were introduced, directly implies dose-mediated gas-phase reactions that alter the cell's gas composition.

Our recently-developed ultrathin $SiN_x$-based gas cell enables the real-time quantification of gaseous species produced during electron illumination at atmospheric pressure *via* EELS due to its low background scattering **(Figure S15)**.[61] Unlike mass spectroscopy-based residual gas analysis, which detects gaseous products throughout the entire TEM cell, EELS provides site-specific analysis without external gas dilution outside the ROI. Moreover, core loss edge fine structures reflect signature bonding modes; the characteristic C *K* edge fine structure peaks for pure $CO_2$ and CO are observed at 289.6 eV and 285.6 eV, respectively. Using time-resolved EELS, we hypothesized, would provide a direct insight into the radiolytic disproportionation of carbon monoxide.

To test this hypothesis, we mixed equivalent partial pressures of $CO_2$ and CO to a total of 760 torr and measured the time-dependent spectral change under continuous electron illumination in TEM mode (27.3 $e^-/Å^2 \cdot s$, $1.23\times10^8$ Gy/s). Initially, two distinct C *K* edge fine structures appeared with a peak height ratio ($h_{CO}/h_{CO2}$) of 0.75. After 20 s of continuous illumination, the CO peak height steadily declined, reducing the ratio to 0.57. After 120 s, this ratio further plummeted to 0.11, suggesting significant CO conversion **(Figure 7B)**. Additional *in situ* EELS measurements on pure CO show that electron radiolysis generates a $CO_2$ peak at 289.6 eV after ~30 s **(Figure 7C)**. Subsequently, both CO and $CO_2$ peak intensities diminished due to increased background thickness from solid carbon deposition. In contrast, identical electron exposure to $CO_2$ produced no detectable changes in peak height or species composition even after 120 s **(Figure 7D)**, with no significant changes in the relative thickness ($t/\lambda_i$) arising from carbon deposition.

Given CO's role as an intermediate in $CO_2$ radiolysis networks, we employed $CO_2$ reaction sets to simulate the $CO/CO_2$ gas mixture's radiolysis behavior. Initial species yields were calculated from the stoichiometric ratio of $CO/CO_2$ (1:1). **Figure 7E** models the temporal evolution of $CO_2$, CO, and C concentration over time in GPTEM with 760 torr $CO/CO_2$. At dose rates below $10^8$ Gy/s, the $CO/CO_2$ ratio remains stable with negligible carbon accumulation. However, above ~$10^8$ Gy/s (easily achievable with a 12 μm$^2$ illumination area and 4.2 nA probe current), the gas composition noticeably changes, and carbon begins to accumulate. At higher doses of $10^{10}$ Gy/s – typical for high-resolution TEM with a 400 nm beam diameter – the $CO/CO_2$ ratio plateaus to a steady state value of ~0.75 within 0.1 s, indicating rapid conversion and equilibration following electron irradiation. However, this model does not incorporate the deposition of gaseous carbon during membrane adsorption or any carbonaceous material deposition, which would drive the equilibrium further to the right. These findings show that interference by the electron beam profoundly affects the dynamic equilibrium of this industrially relevant chemical process. Consequently, electron dose rates must be carefully considered in studies of *in situ* reaction systems involving CO to prevent coking and adventitiously decreased partial pressures.

## Conclusions

Radiation chemistry, while assumed negligible in open-cell E-TEM, plays a nontrivial role in closed-cell E-TEM. From both an experimental and theoretical purview, our findings reveal that, unlike open-cell E-TEM operating at reduced gas pressures, electron radiolysis significantly influences *in situ* reactions within closed-cell gas S/TEM setups at near-atmospheric pressure. Moreover, even at low doses in STEM mode, generated reactive species can accelerate Al nanocube oxidation compared to non-illuminated conditions. We further demonstrate that the very act of electron illumination can fundamentally alter the gas composition, such as inducing carbonaceous material deposition during CO exposure, even at low magnifications. For *in situ* or *operando* experiments involving CO, we recommend the use of STEM mode since the interlacing probe promotes species diffusion and prevents CO consumption in the viewing area, thereby providing a more dynamic and representative observation environment. Collectively, the work established herein provides a robust framework for the high-precision dosimetry of both *in situ* gas and liquid reactions involving catalytic and morphological transformations at the nanoscale. By mapping the intricate interactions between electron radiation and chemical systems, we hope this work provides the reader with nuanced insights into previously obscured physicochemical processes.

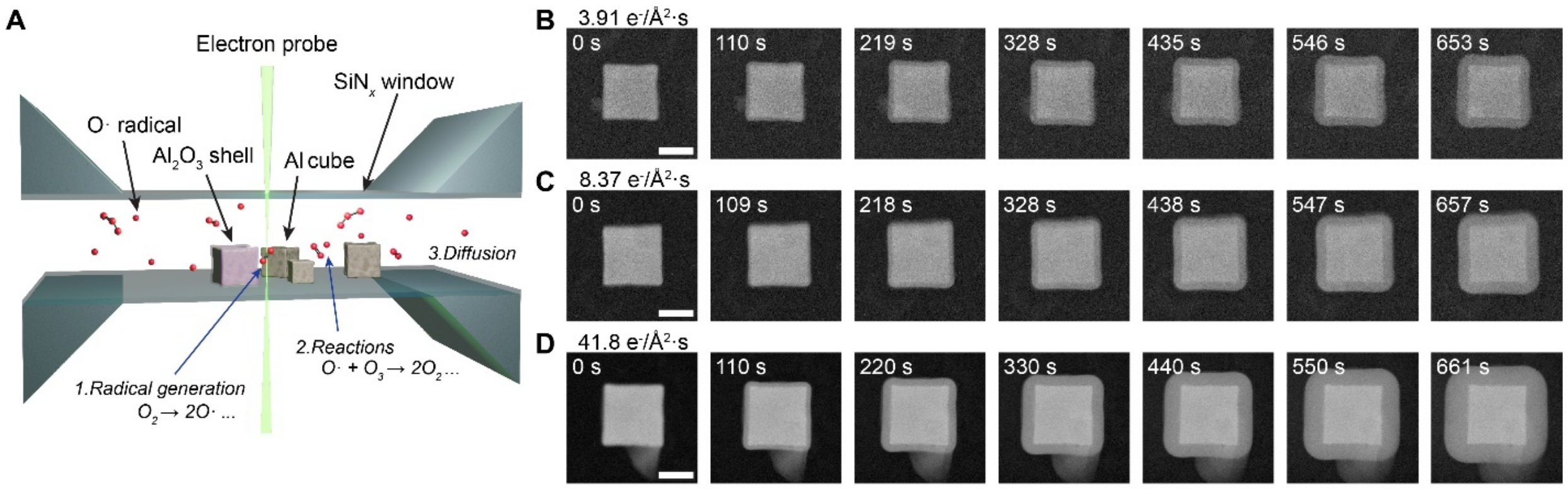

**Figure 1. Electron-dose dependence of reaction kinetics in a high-pressure GPTEM. (A)** Schematic illustration of the SiN$_x$-based closed gas cell S/TEM. **(B-D)** Time-series HAADF-STEM images (excerpted from Movies S5-S7) showing the oxidation behavior of 200 nm aluminum nanocubes with different electron dose rates. 10 torr $O_2$ is introduced at room temperature. Scale bars are 100 nm. The electron dose of the time series is (B) 3.91 e$^-$/Å$^2$·s (1.53 × $10^7$ Gy/s), (C) 8.37 e$^-$/Å$^2$·s (3.28 × $10^7$ Gy/s), and (D) 41.8 e$^-$/Å$^2$·s (1.64 × $10^8$ Gy/s).

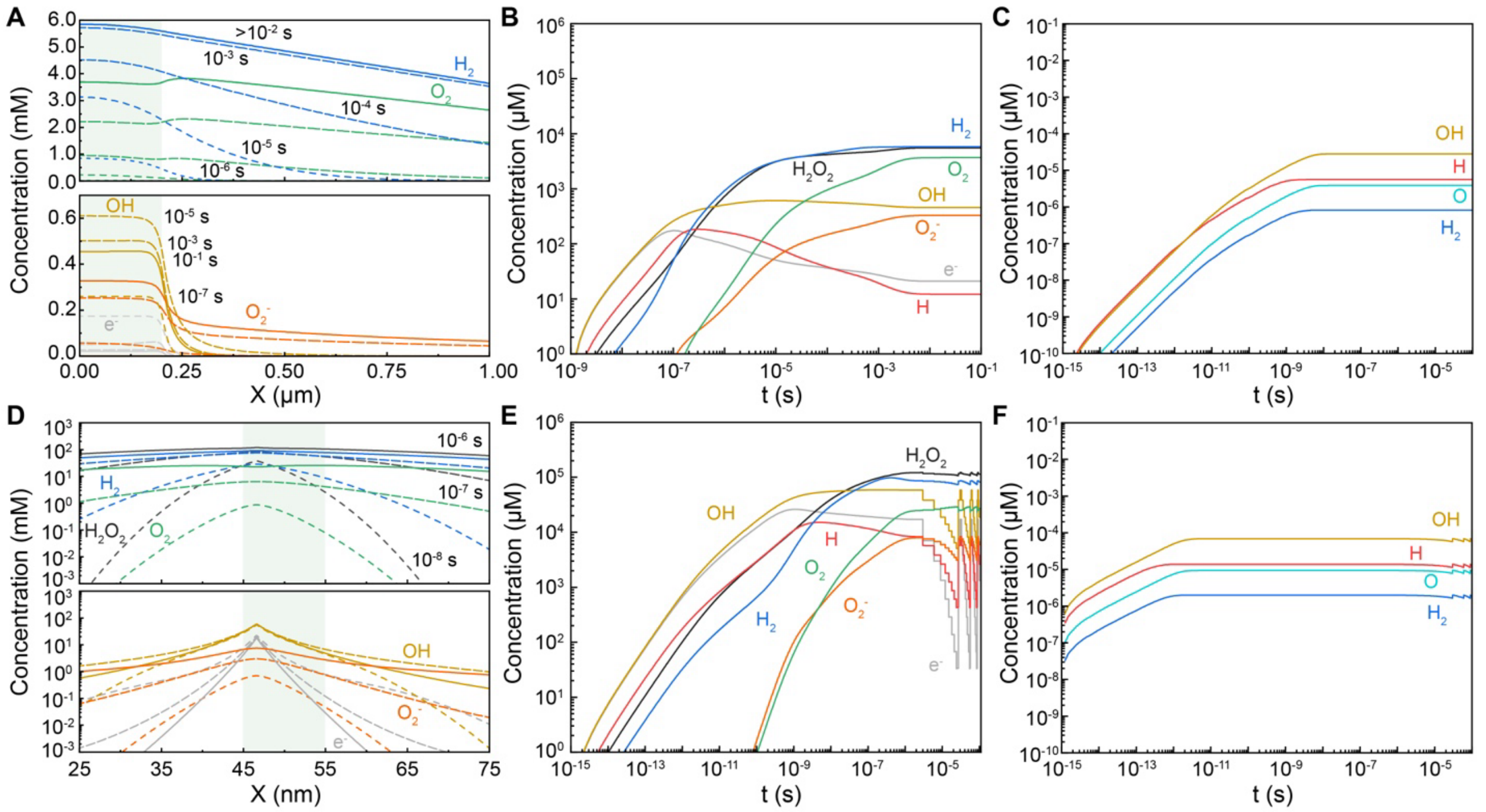

**Figure 2. Calculated early-stage diffusion profiles of radiolysis species and a comparison of radiolysis species concentrations depending on the fluidic phase and illumination modes. (A)** Diffusion profiles of $H_2$, $O_2$, $OH\cdot$, $O_2^-$, and $e_h^-$ in LPTEM (water). The green shaded region is the area illuminated by the electron beam (r = 200 nm), with a dose rate is $9.4 \times 10^9$ Gy/s. **(B)** Timewise evolution of radiolysis species concentrations in liquid water following TEM irradiation. **(C)** Timewise evolution of radiolysis species in 10 torr water vapor with TEM illumination. **(D)** Diffusion profiles of $H_2$, $O_2$, $OH\cdot$, $O_2^-$, and $e_h^-$ in liquid water during initial electron-probe irradiation. In STEM mode, the probe diameter, separation and current are 1 Å, 1 nm, and 62 pA respectively, yielding a dose rate of $6.66 \times 10^7$ Gy/s within the scanning area. **(E)** Timewise evolution of radiolysis species in liquid water in STEM mode. **(F)** Time evolution of radiolysis species in 10 torr water vapor in STEM mode. (E-F) The concentration profile was probed at the center of the initial probe location and the dwell time is 3 µs. The length of the dash in (A, D) indicates different time durations.

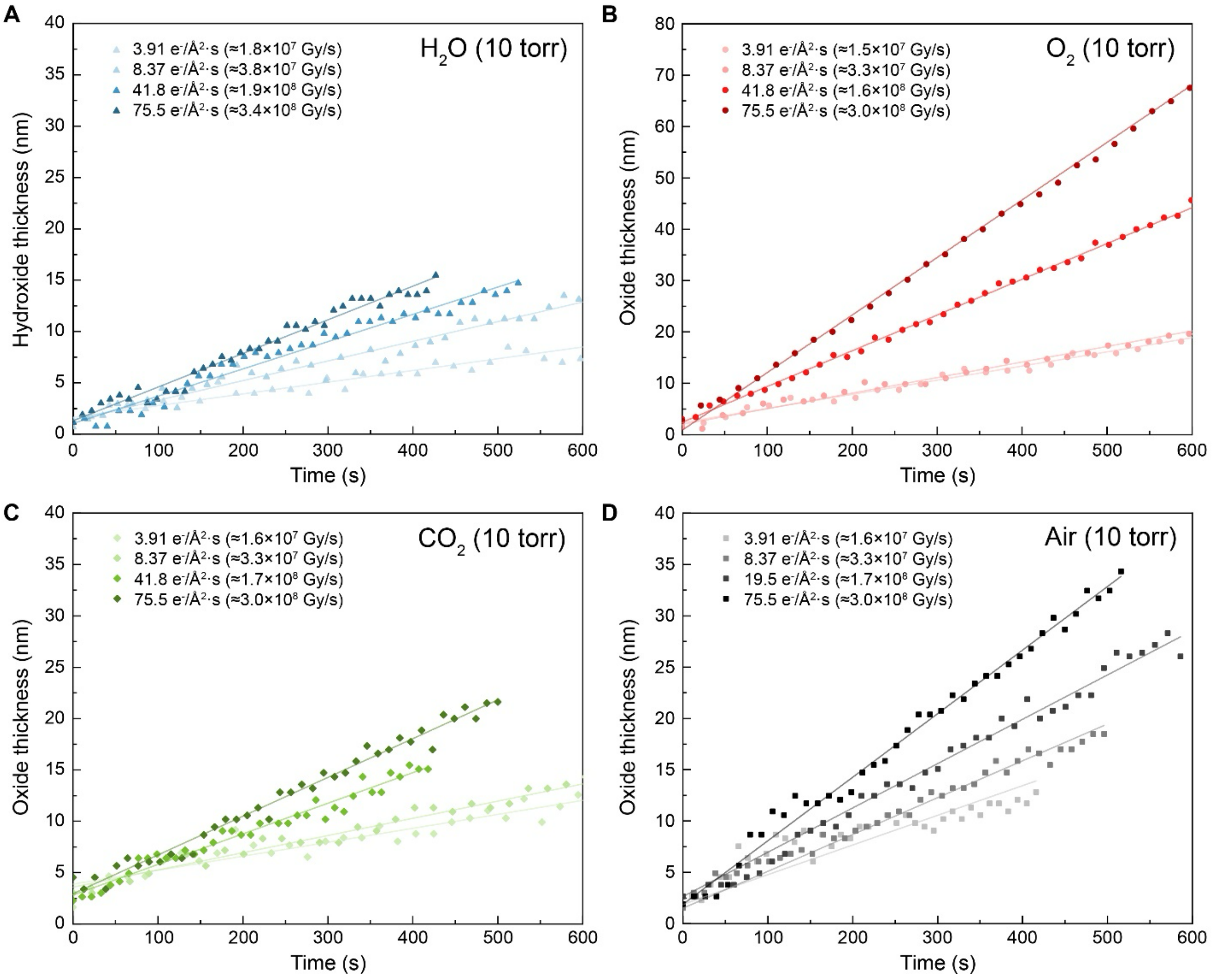

**Figure 3. Reaction of an Al nanocube under different electron doses in STEM with (A)** 10 torr water vapor, **(B)** 10 torr $O_2$, **(C)** 10 torr $CO_2$, and **(D)** 10 torr dry air. (A-D) In Movies S5-S7, a 1 Å-electron probe was scanned over an image area of approximately 500 nm × 500 nm with 512 × 512 pixels.

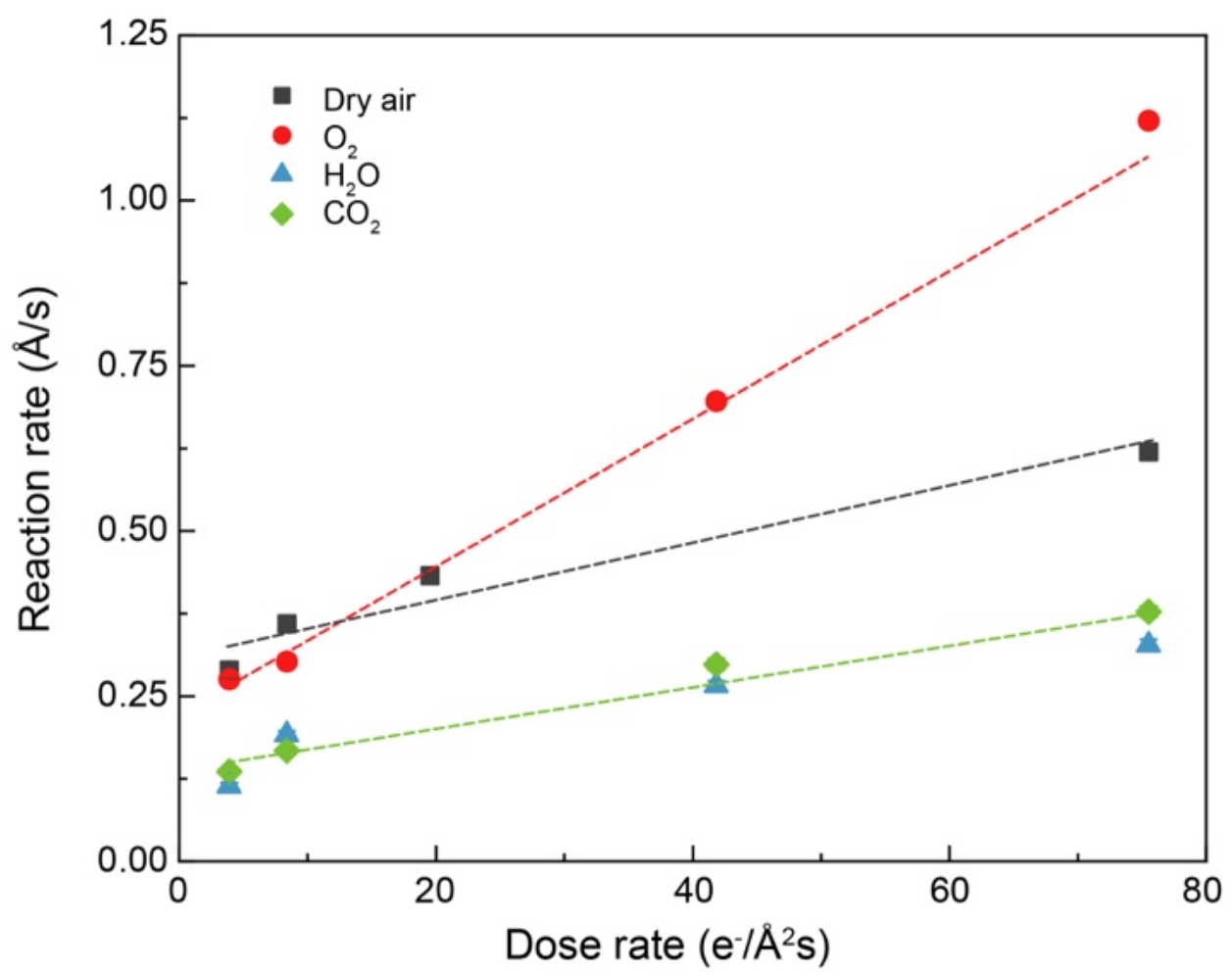

**Figure 4.** Al oxidation rate dependence on electron dose with four different gaseous species at 10 torr. We omit a trendline for oxidation by $H_2O$ to avoid a direct comparison of its reaction kinetics.

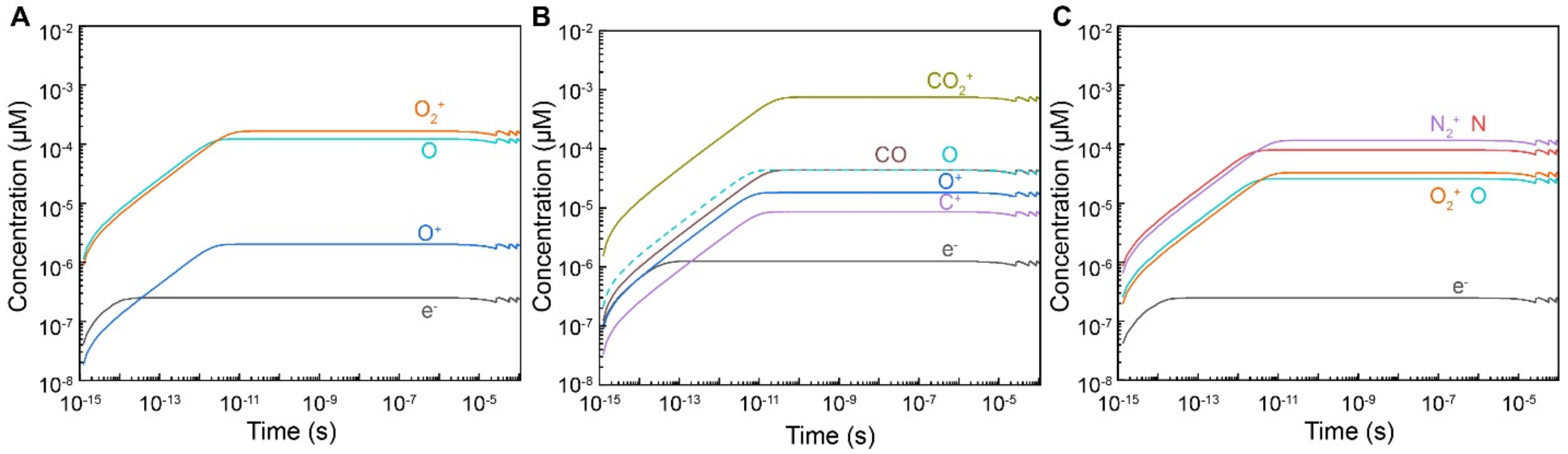

**Figure 5. Calculated major radiolysis products under closed-cell GPSTEM conditions. (A)** 10 torr $O_2$, **(B)** 10 torr $CO_2$, and **(C)** 10 torr dry air. The probe diameter, step and current are 1 Å, 1 nm and 62 pA respectively, yielding a dose rate of $5.8 \times 10^7$ Gy/s. The concentration profiles are probed at the center of the initial probe location, and the dwell time is 3 μs.

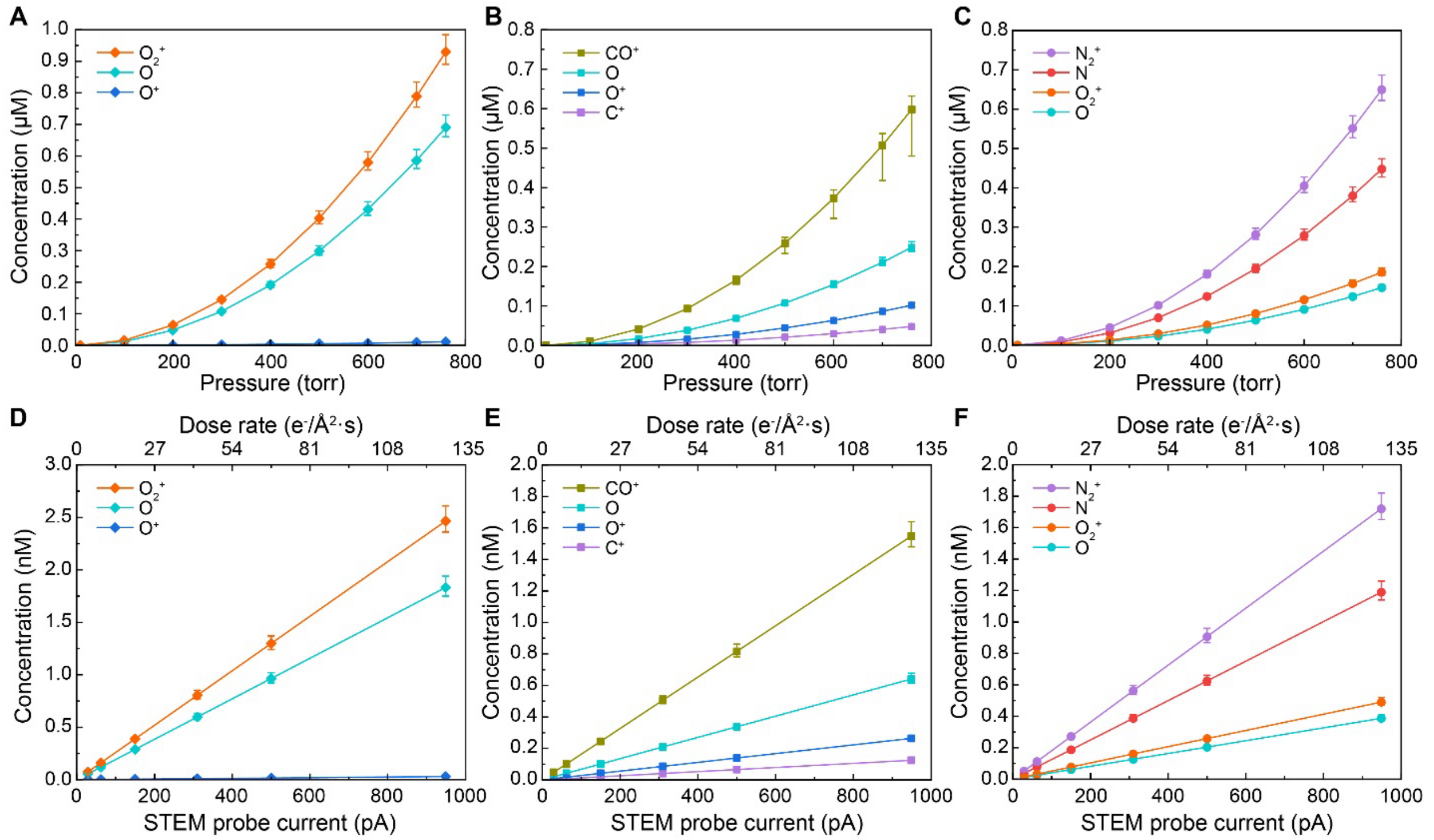

**Figure 6. Calculated steady-state concentrations of major radiolysis products in STEM mode under different conditions**. **(A-C)** variable pressures and fixed dose rate at 8.37 $e^{-}/Å^{2}\cdot s$ (~3.1 × $10^{7}$ Gy/s), and **(D-F)** variable electron dose rates and fixed pressure at 10 torr. The error bars indicate the highest and the lowest concentrations due to variations in probe position. In (A, D), (B, E), and (C, F), the gas is $O_2$, $CO_2$ and dry air, respectively.

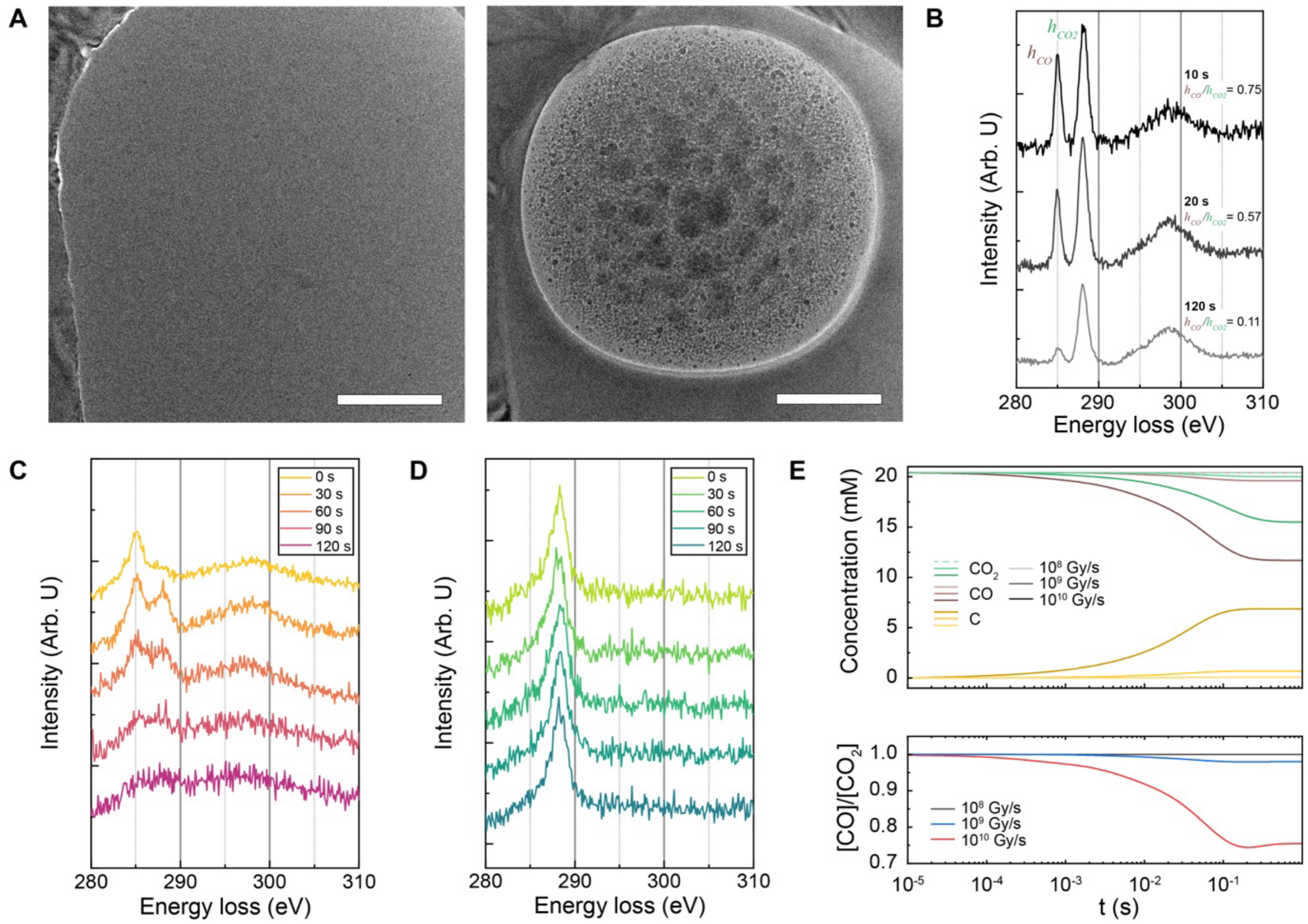

**Figure 7. Radiation chemistry within pure $CO_2$ and mixed $CO/CO_2$. (A)** Deposition of carbonaceous materials after continuous illumination with a dose rate of ~$1.2 \times 10^8$ Gy/s in the $CO_2$-filled (left) and $CO/CO_2$ gas cell (right). The scale bar is 1 μm. **(B)** Timewise evolution of EELS profiles of a 50% CO and 50% $CO_2$ mixed gas cell with a total pressure of 760 torr. Timewise evolution of EELS profiles of **(C)** pure CO- and **(D)** pure $CO_2$-filled gas cells at 760 torr. **(E)** Simulated timewise evolution of CO, $CO_2$, and C concentrations within the mixed $CO/CO_2$ gas cell per different dose rates (top) and the CO:$CO_2$ ratio (bottom).

| | | $\alpha$ (at 62 pA) | $\beta$ (at 10 torr) |
|---|---|---|---|
| Oxygen ($O_2$) | $O_2^+$ | $1.61\times10^{-6}$ | $2.60\times10^{-6}$ |
| | O | $1.19\times10^{-6}$ | $1.93\times10^{-6}$ |
| | $O^+$ | $1.96\times10^{-8}$ | $3.16\times10^{-6}$ |
| | $e^-$ | $2.44\times10^{-9}$ | $3.92\times10^{-6}$ |
| | | | |
| Carbon dioxide ($CO_2$) | $CO^+$ | $1.04\times10^{-6}$ | $1.63\times10^{-6}$ |
| | O | $4.27\times10^{-7}$ | $6.75\times10^{-7}$ |
| | $O^+$ | $1.77\times10^{-7}$ | $2.78\times10^{-7}$ |
| | $C^+$ | $8.32\times10^{-8}$ | $1.31\times10^{-7}$ |
| | $e^-$ | $1.20\times10^{-8}$ | $1.89\times10^{-8}$ |
| | | | |
| Air (78.5% $N_2$ + 21.5% $O_2$) | $N_2^+$ | $1.13\times10^{-6}$ | $1.81\times10^{-6}$ |
| | N | $7.76\times10^{-7}$ | $1.25\times10^{-6}$ |
| | $O_2^+$ | $3.20\times10^{-7}$ | $5.17\times10^{-7}$ |
| | O | $2.53\times10^{-7}$ | $4.08\times10^{-7}$ |
| | $e^-$ | $2.42\times10^{-9}$ | $3.88\times10^{-9}$ |

**Table 1.** Pressure coefficients ($\alpha$) at 62 pA fixed probe current and current coefficients ($\beta$) at fixed pressure (10 torr).

ASSOCIATED CONTENT

The following files are available free of charge.

Supporting Information (PDF)

Movie S1-7 (MP4)

AUTHOR INFORMATION

**Corresponding Author**

*Xiaobing Hu (email: xbhu@northwestern.edu)

*Vinayak P. Dravid (email: v-dravid@northwestern.edu)

**Author Contributions**

K.K. conducted gas cell experiments and generated code for radical formation. S.S. analyzed movie data. N.S.C. synthesized nanomaterials and co-performed gas cell experiments. X.H. and V.P.D. conceived and coordinated the project. All authors were involved in data analysis and subsequent writing process and have given approval to the final version of the manuscript.

**Funding Sources**

V.P.D., K.K., N.S.C., and C.A.M. acknowledge support from the Air Force Office of Scientific Research (AFOSR) with the grant number of AFOSR FA9550-22-1-0300. N.S.C. and G.C.S. acknowledge funding from Army Research Office Grant No. W911NF-20-1-0105. N.S.C. acknowledges support by the National Science Foundation Graduate Research Fellowship under Grant No. DGE-2234667. This work made use of the EPIC and NUFAB facility of Northwestern University's NU*ANCE* Center, which has received support from the SHyNE Resource (NSF ECCS-2025633), the IIN, and Northwestern's MRSEC program (NSF DMR- 2308691).

ACKNOWLEDGMENT

The authors thank Dr. J. K. H. Orbeck and Dr. S. H. Petrosko (Northwestern University) for providing editorial input.

ABBREVIATIONS

S/TEM, Scanning/Transmission Electron Microscope/Microscopy; E-TEM, Environmental Transmission Electron Microscopy; BSE, Backscattered and Secondary Electron; LP(S)TEM, Liquid Phase (Scanning) Transmission Electron Microscopy; GP(S)TEM, Gas Phase (Scanning) Transmission Electron Microscopy; ROI, Region of interest; IMFP, Inelastic Mean Free Path; PDE, Partial Differential Equation; LHS, Left Hand Side; RHS, Right Hand Side; ODE, Ordinary Differential Equation; EELS, Electron Energy Loss Spectroscopy; EDS, Energy Dispersive X-ray Spectroscopy.

REFERENCES

(1) Egerton, R.; Li, P.; Malac, M. Radiation Damage in the TEM and SEM. *Micron* **2004**, *35* (6), 399-409.

(2) Schneider, N. M.; Norton, M. M.; Mendel, B. J.; Grogan, J. M.; Ross, F. M.; Bau, H. H. Electron–water Interactions and Implications for Liquid Cell Electron Microscopy. *The Journal of Physical Chemistry C* **2014**, *118* (38), 22373-22382.

(3) Plante, I.; Cucinotta, F. A. Cross Sections for the Interactions of 1 eV–100 MeV Electrons in Liquid Water and Application to Monte-Carlo Simulation of HZE Radiation Tracks. *New Journal of Physics* **2009**, *11* (6), 063047.

(4) Zheng, H.; Claridge, S. A.; Minor, A. M.; Alivisatos, A. P.; Dahmen, U. Nanocrystal Diffusion in a Liquid Thin Film Observed by *in situ* Transmission Electron Microscopy. *Nano Letters* **2009**, *9* (6), 2460-2465.

(5) Grogan, J. M.; Schneider, N. M.; Ross, F. M.; Bau, H. H. Bubble and Pattern Formation in Liquid Induced by an Electron Beam. *Nano Letters* **2014**, *14* (1), 359-364.

(6) Cazaux, J. Correlations Between Ionization Radiation Damage and Charging Effects in Transmission Electron Microscopy. *Ultramicroscopy* **1995**, *60* (3), 411-425.

(7) Yuk, J. M.; Seo, H. K.; Choi, J. W.; Lee, J. Y. Anisotropic Lithiation Onset in Silicon Nanoparticle Anode Revealed by *in situ* Graphene Liquid Cell Electron Microscopy. *ACS Nano* **2014**, *8* (7), 7478-7485.

(8) Park, J. Y.; Kim, S. J.; Chang, J. H.; Seo, H. K.; Lee, J. Y.; Yuk, J. M. Atomic Visualization of a Non-Equilibrium Sodiation Pathway in Copper Sulfide. *Nature Communications* **2018**, *9* (1), 922.

(9) Cheong, J. Y.; Chang, J. H.; Seo, H. K.; Yuk, J. M.; Shin, J. W.; Lee, J. Y.; Kim, I.-D. Growth Dynamics of Solid Electrolyte Interphase Layer on $SnO_2$ Nanotubes Realized by Graphene Liquid Cell Electron Microscopy. *Nano Energy* **2016**, *25*, 154-160.

(10) Yuk, J. M.; Park, J.; Ercius, P.; Kim, K.; Hellebusch, D. J.; Crommie, M. F.; Lee, J. Y.; Zettl, A.; Alivisatos, A. P. High-resolution EM of Colloidal Nanocrystal Growth Using Graphene Liquid Cells. *Science* **2012**, *336* (6077), 61-64.

(11) Chen, Q.; Yuk, J. M.; Hauwiller, M. R.; Park, J.; Dae, K. S.; Kim, J. S.; Alivisatos, A. P. Nucleation, Growth, and Superlattice Formation of Nanocrystals Observed in Liquid Cell Transmission Electron Microscopy. *MRS Bulletin* **2020**, *45* (9), 713-726.

(12) Zheng, H.; Smith, R. K.; Jun, Y.-w.; Kisielowski, C.; Dahmen, U.; Alivisatos, A. P. Observation of Single Colloidal Platinum Nanocrystal Growth Trajectories. *Science* **2009**, *324* (5932), 1309-1312.

(13) Kim, S. Y.; Dae, K. S.; Koo, K.; Kim, D.; Park, J.; Yuk, J. M. Sequential Growth and Etching of Gold Nanocrystals Revealed by High-Resolution Liquid Electron Microscopy. *Physica Status Solidi (a)* **2019**, *216* (7), 1800949.

(14) Hauwiller, M. R.; Ye, X.; Jones, M. R.; Chan, C. M.; Calvin, J. J.; Crook, M. F.; Zheng, H.; Alivisatos, A. P. Tracking the Effects of Ligands on Oxidative Etching of Gold Nanorods in Graphene Liquid Cell Electron Microscopy. *ACS Nano* **2020**, *14* (8), 10239-10250.

(15) Sun, M.; Tian, J.; Chen, Q. The Studies on Wet Chemical Etching via *in situ* Liquid Cell TEM. *Ultramicroscopy* **2021**, *231*, 113271.

(16) Wang, X.; Li, Y.; Meng, Y. S. Cryogenic Electron Microscopy for Characterizing and Diagnosing Batteries. *Joule* **2018**, *2* (11), 2225-2234.

(17) Mishyna, M.; Volokh, O.; Danilova, Y.; Gerasimova, N.; Pechnikova, E.; Sokolova, O. Effects of Radiation Damage in Studies of Protein-DNA Complexes by Cryo-EM. *Micron* **2017**, *96*, 57-64.

(18) Grant, T.; Grigorieff, N. Measuring the Optimal Exposure for Single Particle Cryo-EM Using a 2.6 Å Reconstruction of Rotavirus VP6. *elife* **2015**, *4*, e06980.

(19) Cho, H.; Jones, M. R.; Nguyen, S. C.; Hauwiller, M. R.; Zettl, A.; Alivisatos, A. P. The Use of Graphene and its Derivatives for Liquid-Phase Transmission Electron Microscopy of Radiation-Sensitive Specimens. *Nano Letters* **2017**, *17* (1), 414-420.

(20) Korpanty, J.; Gnanasekaran, K.; Venkatramani, C.; Zang, N.; Gianneschi, N. C. Organic Solution-Phase Transmission Electron Microscopy of Copolymer Nanoassembly Morphology and Dynamics. *Cell Reports Physical Science* **2022**, *3* (3).

(21) Korpanty, J.; Parent, L. R.; Gianneschi, N. C. Enhancing and Mitigating Radiolytic Damage to Soft Matter in Aqueous Phase Liquid-Cell Transmission Electron Microscopy in the Presence of Gold Nanoparticle Sensitizers or Isopropanol Scavengers. *Nano Letters* **2021**, *21* (2), 1141-1149.

(22) Koo, K.; Shen, B.; Baik, S.-I.; Mao, Z.; Smeets, P. J.; Cheuk, I.; He, K.; Dos Reis, R.; Huang, L.; Ye, Z. Formation Mechanism of High-Index Faceted Pt-Bi Alloy Nanoparticles by Evaporation-Induced Growth from Metal Salts. *Nature Communications* **2023**, *14* (1), 3790.

(23) He, K.; Kim, K.; Villa, C. J.; Ribet, S. M.; Smeets, P.; Dos Reis, R.; Voorhees, P. W.; Hu, X.; Dravid, V. P. Degeneration Behavior of Cu Nanowires Under Carbon Dioxide Environment: an *in situ/operando* Study. *Nano Letters* **2021**, *21* (16), 6813-6819.

(24) Hutzler, A.; Fritsch, B.; Jank, M. P.; Branscheid, R.; Martens, R. C.; Spiecker, E.; März, M. *In situ* Liquid Cell TEM Studies on Etching and Growth Mechanisms of Gold Nanoparticles at a Solid–Liquid–Gas Interface. *Advanced Materials Interfaces* **2019**, *6* (20), 1901027.

(25) Wang, M.; Park, C.; Woehl, T. J. Quantifying the Nucleation and Growth Kinetics of Electron Beam Nanochemistry with Liquid Cell Scanning Transmission Electron Microscopy. *Chemistry of Materials* **2018**, *30* (21), 7727-7736.

(26) Liu, L.; Sassi, M.; Zhang, X.; Nakouzi, E.; Kovarik, L.; Xue, S.; Jin, B.; Rosso, K. M.; De Yoreo, J. J. Understanding the Mechanisms of Anisotropic Dissolution in Metal Oxides by Applying Radiolysis Simulations to Liquid-Phase TEM. *Proceedings of the National Academy of Sciences* **2023**, *120* (23), e2101243120.

(27) Ambrožič, B.; Prašnikar, A.; Hodnik, N.; Kostevšek, N.; Likozar, B.; Rožman, K. Ž.; Šturm, S. Controlling the Radical-Induced Redox Chemistry Inside a Liquid-Cell TEM. *Chemical Science* **2019**, *10* (38), 8735-8743.

(28) Crook, M. F.; Moreno-Hernandez, I. A.; Ondry, J. C.; Ciston, J.; Bustillo, K. C.; Vargas, A.; Alivisatos, A. P. EELS Studies of Cerium Electrolyte Reveal Substantial Solute Concentration Effects in Graphene Liquid Cells. *Journal of the American Chemical Society* **2023**, *145* (12), 6648-6657.

(29) Lee, S.; Schneider, N. M.; Tan, S. F.; Ross, F. M. Temperature Dependent Nanochemistry and Growth Kinetics Using Liquid Cell Transmission Electron Microscopy. *ACS Nano* **2023**, *17* (6), 5609-5619.

(30) Merkens, S.; De Salvo, G.; Chuvilin, A. The Effect of Flow on Radiolysis in Liquid Phase-TEM Flow Cells. *Nano Express* **2023**, *3* (4), 045006.

(31) Chen, X.; Shan, W.; Wu, D.; Patel, S. B.; Cai, N.; Li, C.; Ye, S.; Liu, Z.; Hwang, S.; Zakharov, D. N. Atomistic Mechanisms of Water Vapor–Induced Surface Passivation. *Science Advances* **2023**, *9* (44), eadh5565.

(32) Nguyen, L.; Hashimoto, T.; Zakharov, D. N.; Stach, E. A.; Rooney, A. P.; Berkels, B.; Thompson, G. E.; Haigh, S. J.; Burnett, T. L. Atomic-Scale Insights into the Oxidation of Aluminum. *ACS Applied Materials & Interfaces* **2018**, *10* (3), 2230-2235.

(33) Wagner, J. B.; Cavalca, F.; Damsgaard, C. D.; Duchstein, L. D.; Hansen, T. W. Exploring the Environmental Transmission Electron Microscope. *Micron* **2012**, *43* (11), 1169-1175.

(34) Koo, K.; Ribet, S. M.; Zhang, C.; Smeets, P. J.; Dos Reis, R.; Hu, X.; Dravid, V. P. Effects of the Encapsulation Membrane in *operando* Scanning Transmission Electron Microscopy. *Nano Letters* **2022**, *22* (10), 4137-4144.

(35) LaVerne, J. A. Radiation Chemistry: Yields of Chemical Species.

(36) Egerton, R. F. *Electron Energy-Loss Spectroscopy in the Electron Microscope*; Springer Science & Business Media, 2011.

(37) Egerton, R. Dose Measurement in the TEM and STEM. *Ultramicroscopy* **2021**, *229*, 113363.

(38) Graupner, A.; Eide, D. M.; Brede, D. A.; Ellender, M.; Lindbo Hansen, E.; Oughton, D. H.; Bouffler, S. D.; Brunborg, G.; Olsen, A. K. Genotoxic Effects of High Dose Rate X-ray and Low Dose Rate Gamma Radiation in ApcMin/+ Mice. *Environmental and Molecular Mutagenesis* **2017**, *58* (8), 560-569.

(39) Sousa, R. Dose Rate Influence on Deep Dose Deposition Using a 6 MV X-Ray Beam from a Linear Accelerator. *Brazilian Journal of Physics* **2009**, *39*, 292-296.

(40) Mehta, K.; Girzikowsky, R. IAEA High-Dose Intercomparison in $^{60}$Co Field. *Applied Radiation and Isotopes* **2000**, *52* (5), 1179-1184.

(41) Liu, C.-J.; Wang, C.-H.; Wang, C.-L.; Hwu, Y.; Lin, C.-Y.; Margaritondo, G. Simple Dose Rate Measurements for a Very High Synchrotron X-ray Flux. *Journal of Synchrotron Radiation* **2009**, *16* (3), 395-397.

(42) Bazalova-Carter, M.; Esplen, N. On the Capabilities of Conventional X-Ray Tubes to Deliver Ultra-High (FLASH) Dose Rates. *Medical Physics* **2019**, *46* (12), 5690-5695.

(43) Fournier, P.; Crosbie, J. C.; Cornelius, I.; Berkvens, P.; Donzelli, M.; Clavel, A.; Rosenfeld, A. B.; Petasecca, M.; Lerch, M. L.; Bräuer-Krisch, E. Absorbed Dose-to-Water Protocol Applied to Synchrotron-Generated X-Rays at Very High Dose Rates. *Physics in Medicine & Biology* **2016**, *61* (14), N349.

(44) Fuller, E. N.; Schettler, P. D.; Giddings, J. C. New Method for Prediction of Binary Gas-Phase Diffusion Coefficients. *Industrial & Engineering Chemistry* **1966**, *58* (5), 18-27.

(45) Langenberg, S.; Carstens, T.; Hupperich, D.; Schweighoefer, S.; Schurath, U. Determination of Binary Gas-Phase Diffusion Coefficients of Unstable and Adsorbing Atmospheric Trace Gases at Low Temperature–Arrested Flow and Twin Tube Method. *Atmospheric Chemistry and Physics* **2020**, *20* (6), 3669-3682.

(46) Tang, M.; Cox, R.; Kalberer, M. Compilation and Evaluation of Gas Phase Diffusion Coefficients of Reactive Trace Gases in the Atmosphere: Volume 1. Inorganic Compounds. *Atmospheric Chemistry and Physics* **2014**, *14* (17), 9233-9247.

(47) Hill, M.; Smith, F. Calculation of Initial and Primary Yields in the Radiolysis of Water. *Radiation Physics and Chemistry* **1994**, *43* (3), 265-280.

(48) *Solve 1-D Parabolic and Elliptic PDEs - MATLAB PDEPE*. The MathWorks Inc., 2023. https://www.mathworks.com/help/matlab/ref/pdepe.html (accessed 2024/01/06).

(49) Koo, K. *kunmo-koo/Gas_radiolysis*. 2024. https://github.com/kunmo-koo/Gas_radiolysis (accessed 2024/01/07).

(50) Evertsson, J.; Bertram, F.; Zhang, F.; Rullik, L.; Merte, L.; Shipilin, M.; Soldemo, M.; Ahmadi, S.; Vinogradov, N.; Carlà, F. The Thickness of Native Oxides on Aluminum Alloys and Single Crystals. *Applied Surface Science* **2015**, *349*, 826-832.

(51) Santos, L. F. M. M. F.; Abrego, F. C.; Torres, K. F. A.; Raimundo, D. S. Inducing Aluminum Oxide Growth at Room Temperature and Atmospheric Pressure Through Low Dose Gamma-Ray Irradiation. *Radiation Physics and Chemistry* **2023**, *204*, 110666.

(52) Vargel, C. *Corrosion of aluminium*; Elsevier, 2020.

(53) Renard, D.; Tian, S.; Ahmadivand, A.; DeSantis, C. J.; Clark, B. D.; Nordlander, P.; Halas, N. J. Polydopamine-Stabilized Aluminum Nanocrystals: Aqueous Stability and Benzo [a] pyrene Detection. *ACS Nano* **2019**, *13* (3), 3117-3124.

(54) Willis, C.; Boyd, A.; Young, M.; Armstrong, D. Radiation Chemistry of Gaseous Oxygen: Experimental and Calculated Yields. *Canadian Journal of Chemistry* **1970**, *48* (10), 1505-1514.

(55) Kummler, R.; Leffert, C.; Im, K.; Piccirelli, R.; Kevan, L.; Willis, C. A Numerical Model of Carbon Dioxide Radiolysis. *The Journal of Physical Chemistry* **1977**, *81* (25), 2451-2463.

(56) Harteck, P.; Dondes, S. Radiation Chemistry of the Fixation of Nitrogen. *Science* **1964**, *146* (3640), 30-35.

(57) Willis, C.; Boyd, A.; Young, M. Radiolysis of Air and Nitrogen–Oxygen Mixtures with Intense Electron Pulses: Determination of a Mechanism by Comparison of Measured and Computed Yields. *Canadian Journal of Chemistry* **1970**, *48* (10), 1515-1525.

(58) Kramida, A. R., Yu.; Reader, J.; NIST ASD Team. *NIST Atomic Spectra Database (version 5.11)*. National Institute of Standards and Technology, 2023. (accessed 2024/01/06).

(59) Rehn, S. M.; Jones, M. R. New Strategies for Probing Energy Systems with *in situ* Liquid-Phase Transmission Electron Microscopy. *ACS Energy Letters* **2018**, *3* (6), 1269-1278.

(60) Dondes, S.; Harteck, P.; Von Weyssenhoff, H. The Gamma Radiolysis of Carbon Monoxide in the Presence of Rare Gases. *Zeitschrift für Naturforschung A* **1964**, *19* (1), 13-18.

(61) Koo, K.; Li, Z.; Liu, Y.; Ribet, S. M.; Fu, X.; Jia, Y.; Chen, X.; Shekhawat, G.; Smeets, P. J. M.; dos Reis, R.; et al. Ultrathin Silicon Nitride Microchip for *in situ/operando* Microscopy with High Spatial Resolution and Spectral Visibility. *Science Advances* **2024**, *10* (3), eadj6417.

Supporting Information

# Electron-Induced Radiation Chemistry in Environmental Transmission Electron Microscopy

*Kunmo Koo, Nikhil S. Chellam, Sangyoon Shim, Chad A. Mirkin, George C. Schatz, Xiaobing Hu*, and Vinayak P. Dravid**

*Xiaobing Hu (email: xbhu@northwestern.edu)

*Vinayak P. Dravid (email: v-dravid@northwestern.edu)

**This PDF file includes:**

- Materials and Methods
- Supplementary Texts I-V
- Figs. S1 to S14
- Supplementary References

**Other Supplementary Information for this manuscript include the following:**

- Movies S1 to S7

Materials and Methods

*Synthetic Procedures*

All commercially available chemicals and reagents were purchased from Sigma-Aldrich and used as received without further purification, unless otherwise stated. A 50 mM stock solution of bis($\eta^5$-cyclopentadienyl)titanium(III) chloride was prepared by reducing 245 mg of bis($\eta^5$-cyclopentadienyl)titanium(IV) chloride with Mn dust (55 mg) in anhydrous tetrahydrofuran (THF) with subsequent separation through a syringe filter. All glassware used under air- and moisture-free conditions were dried at 180 ºC overnight before usage.

*Synthesis of Al Nanocubes:*[1]

*(CAUTION: Aluminum hydride violently reacts with water and organic halides.)* Syntheses of Al nanoparticles were performed using previously reported methods with slight modifications. Syntheses were undertaken either using typical Schlenk techniques or in a nitrogen-filled glovebox containing <1 ppm $O_2$ and <2 ppm $H_2O$. In a typical synthesis of ≈200 nm nanocubes, dimethylethylamine alane (1 mL, 0.5 M in toluene) was added to anhydrous 1,2-dimethoxyethane (9 mL) and heated to 80 ºC under vigorous stirring. Next, bis($\eta^5$- cyclopentadienyl)titanium(III) chloride (100 μL, 50 mM in THF) was added. The reaction mixture immediately turned a pale lavender in color, became golden-brown after 15 minutes, then turned opaque grey as particles began to form, then brown as particles grew larger. After approximately 2 hours, the reaction was quenched by slowly adding dibutyl phosphate (1 mL, 1% v/v in toluene). After bubbling ceased, the reaction vessel was exposed to ambient conditions, centrifuged (2000 ×g/15 min) twice in toluene (2 × 20 mL), and redispersed in acetonitrile (10 mL).

*Stripping of Nanoparticle Native Ligand Shell:*[2]

*(CAUTION: Triethyloxonium tetrafluoroborate is a potent alkylating agent. While not volatile, care must be taken to minimize ingestion or contact with skin.)* Under a $N_2$ atmosphere, triethyloxonium tetrafluoroborate (750 μL, 1 M in dichloromethane) was added to a particle suspension in acetonitrile (10 mL). The mixture was sonicated for 10 minutes, during which the particles precipitated out of solution. The mixture was then exposed to ambient conditions to

decant the supernatant, centrifuged twice with acetonitrile (2000 ×g/5 min), and finally re-suspended in hexanes (10 mL).

*S/TEM Characterization*

Both conventional transmission electron microscopy (TEM) and *in-situ* gas cell scanning TEM (STEM) analyses were carried out on a JEOL ARM 200CF probe-corrected microscope under 200 keV. This microscope was equipped with two silicon drift detectors (SDDs) and a Gatan Quantum electron energy loss (EEL) spectrometer. The area of a single SDD is 100 $mm^2$, and the total solid angle for energy dispersive spectrum (EDS) collection is approximately 1.7 sr. The EEL spectrum (EELS) was acquired with a Gatan K2 direct electron detector. The convergence angle for STEM analysis was 28 mrad. The collection angle of high-angle annular dark field (HAADF) imaging was 90-250 mrad. The outer collection angle for EELS analysis was approximately 74 mrad. The energy dispersion for EELS acquisition was 0.1 eV/Ch and the energy resolution was 0.6 eV.

*In situ STEM experiments*

The oxidation of Al nanocubes and the EELS measurements of the compositional changes in mixed CO-$CO_2$ gas were conducted in a Protochips Atmosphere 210 atmosphere system. The closed gas cell holder can accommodate two $SiN_x$-membrane windowed microchips for environmental encapsulation. For the oxidation experiments, microchips with ceramic heating membranes were used to reduce contamination and surface-adsorbed oxidative species *via* heating in a reducing environment without electron beam illumination (100 torr $H_2$ at 150 ºC). The thickness of the native oxide layer did not change after annealing. The *in situ* STEM movies tracking the oxidation process were taken with the Protochips AXON software with drift-correction. For the EELS measurements, custom-made ultrathin (UT) $SiN_x$ gas chips comprising two 9 nm membranes were utilized to significantly reduce plural electron scattering from the encapsulation membrane ($t/\lambda_i \approx 0.3$).[3]

*Physical model*

The MATLAB scripts and custom functions are available on GitHub.[4] A detailed theoretical background and an explanation of the code are available in Supplementary Text I-IV. The matrix dimensions used for the partial differential equation (PDE) are defined by the number of species, 1-D diffusion mesh size (1000), and integration timestep ($N \times x \times t$). The maximum timestep was determined to be approximately $3 \times 10^5$, yielding a total size of 30-40 GB of 8-byte double-precision floating point numbers. The timestep consideration was based on limitations of physical memory that a typical desktop computer can handle in the year 2024. To maintain integration accuracy, longer-duration calculations were divided into several runs and stitched together.

Supplementary Text I: *G*-values and stopping power

The *G*-value is a coefficient which describes the number of generated species per 100 eV of deposited energy.[5] We adopted the *G*-value for liquid $H_2O$ from Schneider *et al*[6] since the *G*-value at 200 keV and 300 keV of energy is nearly identical. A list of the *G*-values used for the numerical models of $H_2O$ vapor, $O_2$[8], $CO_2$[9], air[8, 10], and mixtures of CO-$CO_2$[11] are provided in Table S1. Note that for the gas mixtures, the *G*-values are estimated from their stoichiometric ratios.

| $H_2O$ (vapor)[16] | | $H_2O$ (liquid)[6] | | $O_2$[8] | | $CO_2$[9] | | Air[8, 10] | | $CO_2$+CO (1:1)[9, 11] | |
|---|---|---|---|---|---|---|---|---|---|---|---|
| $H_2$ | 0.5 | $e_h^-$ | 3.47 | $e^-$ | 3.27 | $e^-$ | 2.96 | $e^-$ | 2.94 | $e^-$ | 1.48 |
| H | 7.4 | H | 1.00 | O | 6.1 | $C^+$ | 0.07 | N | 4.68 | C | 3.7 |
| OH | 6.3 | $H_2$ | 0.17 | $O^+$ | 0.1 | O | 0.51 | $N_2$ | -4.57 | $C^+$ | 0.04 |
| O | 1.1 | OH | 3.63 | $O_2$ | -6.32 | $O^+$ | 0.21 | $N_2^+$ | 2.15 | O | 0.26 |
| $H_2O$ | -7.4 | $H_2O_2$ | 0.47 | $O_2^+$ | 3.17 | CO | 0.21 | O | 1.28 | $O^+$ | 0.11 |
| | | $HO_2$ | 0.08 | | | $CO^+$ | 0.51 | $O_2$ | -1.33 | CO | -4.54 |
| | | $H_3O^+$ | 4.42 | | | $CO_2$ | -3.03 | $O_2^+$ | 0.67 | $CO^+$ | 0.26 |
| | | $OH^-$ | 0.95 | | | $CO_2^+$ | 2.24 | | | $CO_2$ | -0.57 |
| | | $H_2O$ | -5.68 | | | | | | | $CO_2^+$ | 1.12 |

**Table S1.** List of *G*-values for $H_2O$ (liquid[6], vapor[16]), $O_2$[8], $CO_2$[9], Air[8, 10], and CO/$CO_2$ (1:1)[9,11].

The electron stopping power *S* [MeV $cm^2/g \cdot e^-$] is adopted from the ESTAR database from NIST.[12] The *S* values for 200 keV electrons are listed in Table S2. The stopping power of CO-$CO_2$ mixed gas is calculated from the average density and molar ratio (0.0016 $g/cm^3$ at STP).

| | *S* [MeV $cm^2/g$] |
|---|---|
| $H_2O$ (vapor) | 2.814 |
| $H_2O$ (liquid) | 2.798 |
| $O_2$ | 2.446 |
| $CO_2$ | 2.480 |
| Air | 2.474 |
| CO-$CO_2$ | 2.475 |
| CO | 2.482 |

**Table S2.** Electron stopping powers *S* per molecule.[12]

Supplementary Text II: Estimation of Diffusion Coefficients

The diffusion coefficients for molecules in a gaseous medium are estimated from the following empirical formula (Equation S1) from Fuller *et al*:[13-15]

$$D_{AB} = \frac{0.00143T^{1.75}}{\sqrt{M_{AB}}\left(V_A^{\frac{1}{3}} + V_B^{\frac{1}{3}}\right)^2 p} \qquad (S1)$$

where $D_{AB}$ ($cm^2 \cdot s^{-1}$) is the diffusivity of species A in gas B at temperature $T$ (K) and pressure $p$ (atm). $M_{AB}$ ($g \cdot mol^{-1}$) and diffusion volume $V$ (dimensionless) are defined in Equations S2-3:

$$M_{AB} = \frac{2}{M_A^{-1} + M_B^{-1}} \qquad (S2)$$

$$V = \sum n_i V_i \qquad (S3)$$

$M_A$ and $M_B$ are the molecular masses (i.e. 1.008 a.m.u. for H) and $5.49 \times 10^{-4}$ a.m.u. is used for the electron mass $M_{e-}$. The diffusion volume is an empirical parameter and can either be collected from the literature or *via* induced summation of individual atoms. The list of diffusion volumes used for the physical model is given in Table S3.

| $H_2O$ (*g*) | | $O_2$ (*g*) | | $CO_2$ (*g*) | | Air (*g*) | |
|---|---|---|---|---|---|---|---|
| $e^-$ | 0 | $e^-$ | 0 | $e^-$ | 0 | $e^-$ | 0 |
| $H^+$ (proton) | 0 | O | 6.11 | C | 15.9 | N | 4.54 |
| $OH^-$ | 8.42 | $O^+$ | 6.11 | $C^+$ | 15.9 | $N^+$ | 4.54 |
| $H_2O_2$ | 19.21 | $O_2$ | 16.3 | O | 6.11 | $N_2$ | 18.5 |
| $HO_2^-$ | 18.61 | $O_2^+$ | 16.3 | $O^+$ | 6.11 | $N_2^+$ | 18.5 |
| H | 2.31 | $O_2^-$ | 16.3 | $O_2$ | 16.3 | $N_4^+$ | 37 |
| OH | 8.42 | $O_3$ | 22.4 | $O_2^+$ | 16.3 | O | 6.11 |
| $O^-$ | 6.11 | $O_3^-$ | 22.4 | CO | 22.01 | $O^+$ | 6.11 |
| $HO_2$ | 18.61 | $O_4^+$ | 28.5 | $CO^+$ | 22.01 | $O_2$ | 16.3 |
| $O_2^-$ | 16.3 | $O_4^-$ | 28.5 | $CO_2$ | 32.2 | $O_2^+$ | 16.3 |
| $O_2$ | 16.3 | | | $CO_2^+$ | 32.2 | $O_2^-$ | 16.3 |
| $H_2$ | 6.12 | | | $CO_3^+$ | 38.31 | $O_3$ | 22.41 |
| $O_3^-$ | 22.41 | | | $CO_4^+$ | 48.5 | $O_3^-$ | 22.41 |
| $O_3$ | 22.41 | | | $CO_4^-$ | 48.5 | NO | 10.65 |
| $HO_3$ | 24.72 | | | $C_2O_2^+$ | 48.1 | $NO^+$ | 10.65 |
| $H_2O$ | 13.1 | | | $C_2O_3^+$ | 54.21 | $NO^-$ | 10.65 |
| | | | | $C_2O_4^+$ | 64.4 | $NO_2$ | 20.84 |
| | | | | $C_3O_4^+$ | 80.3 | $NO_2^-$ | 20.84 |
| | | | | | | $NO_3^-$ | 26.95 |
| | | | | | | $N_2O$ | 24.61 |

**Table S3.** List of diffusion volume (*V*) of species used for this model.

Note that we have disregarded the effect of charge towards volume in this calculation model; thus the diffusion volume of electrons ($e^-$) and protons ($H^+$) are assumed to be zero.

Supplementary Text III: Kinetic models

The kinetic model for water is excerpted from Schneider *et al.*[6]

| Equilibria | $pK_a$ |
|---|---|
| $H_2O \leftrightarrow H^+ + OH^-$ | 13.999 |
| $H_2O_2 \leftrightarrow H^+ + HO_2^-$ | 11.65 |
| $OH \leftrightarrow H^+ + O^-$ | 11.9 |
| $HO_2 \leftrightarrow H^+ + O_2^-$ | 4.57 |
| $H \leftrightarrow H^+ + e^-_{aq}$ | 9.77 |
| Chemical Reaction | Rate constant [$M^{-1} \cdot s^{-1}$ otherwise noted] |
| $H^+ + OH^- \rightarrow H_2O$ | $1.4 \times 10^{11}$ |
| $H_2O \rightarrow H^+ + OH^-$ | $k_7 \times K_2$ / $[H_2O]$ $s^{-1}$ |
| $H_2O_2 \rightarrow H^+ + HO_2^-$ | $k_{10} \times K_3$ $s^{-1}$ |
| $H^+ + HO_2^- \rightarrow H_2O_2$ | $5.0 \times 10^{10}$ |
| $H_2O_2 + OH^- \rightarrow HO_2^- + H_2O$ | $1.3 \times 10^{10}$ |
| $HO_2^- + H_2O \rightarrow H_2O_2 + OH^-$ | $k_{11} \times K_2$ / $K_3[H_2O]$ |
| $e_{aq}^- + H_2O \rightarrow H + OH^-$ | $1.9 \times 10^1$ |
| $H + OH^- \rightarrow e_{aq}^- + H_2O$ | $2.2 \times 10^7$ |
| $H \rightarrow e_{aq}^- + H^+$ | $k_{16} \times K_6$ $s^{-1}$ |
| $e_{aq}^- + H^+ \rightarrow H$ | $2.3 \times 10^{10}$ |
| $OH + OH^- \rightarrow O^- + H_2O$ | $1.3 \times 10^{10}$ |
| $O^- + H_2O \rightarrow OH + OH^-$ | $k_{17} \times K_2$ / $K_4[H_2O]$ |
| $OH \rightarrow O^- + H^+$ | $k_{20} \times K_4$ $s^{-1}$ |
| $O^- + H^+ \rightarrow OH$ | $1.0 \times 10^{11}$ |
| $HO_2 \rightarrow O_2^- + H^+$ | $k_{22} \times K_5$ $s^{-1}$ |
| $O_2^- + H^+ \rightarrow HO_2$ | $5.0 \times 10^{10}$ |
| $HO_2 + OH^- \rightarrow O_2^- + H_2O$ | $5.0 \times 10^{10}$ |
| $O_2^- + H_2O \rightarrow HO_2 + OH^-$ | $k_{23} \times K_2$ / $K_5[H_2O]$ |
| $e_{aq}^- + OH \rightarrow OH^-$ | $3.0 \times 10^{10}$ |
| $e_{aq}^- + H_2O_2 \rightarrow OH + OH^-$ | $1.1 \times 10^{10}$ |
| $e_{aq}^- + O_2^- + H_2O \rightarrow HO_2^- + OH^-$ | $1.3 \times 10^{10}$ / $[H_2O]$ $M^{-2}$ $s^{-1}$ |
| $e_{aq}^- + HO_2 \rightarrow HO_2^-$ | $2.0 \times 10^{10}$ |
| $e_{aq}^- + O_2 \rightarrow O_2^-$ | $1.9 \times 10^{10}$ |
| $e_{aq}^- + e_{aq}^- + 2H_2O \rightarrow H_2 + 2OH^-$ | $5.5 \times 10^9$ / $[H_2O]^2$ $M^{-3}$ $s^{-1}$ |
| $e_{aq}^- + H + H_2O \rightarrow H_2 + OH^-$ | $2.5 \times 10^{10}$ / $[H_2O]$ $M^{-2}$ $s^{-1}$ |
| $e_{aq}^- + HO_2 \rightarrow O^- + OH^-$ | $3.5 \times 10^9$ |
| $e_{aq}^- + O^- + H_2O \rightarrow OH^- + OH^-$ | $2.2 \times 10^{10}$ / [$H_2O$] $M^{-2}$ $s^{-1}$ |
| $e_{aq}^- + O_3^- + H_2O \rightarrow O_2 + OH^- + OH^-$ | $1.6 \times 10^{10}$ / [$H_2O$] $M^{-2}$ $s^{-1}$ |
| $e_{aq}^- + O_3 \rightarrow O_3^-$ | $3.6 \times 10^{10}$ |
| $H + H_2O \rightarrow H_2 + OH$ | $1.1 \times 10^1$ |
| $H + O^- \rightarrow OH^-$ | $1.0 \times 10^{10}$ |
| $H + HO_2^- \rightarrow OH + OH^-$ | $9.0 \times 10^7$ |
| $H + O_3^- \rightarrow OH^- + O_2$ | $1.0 \times 10^{10}$ |
| $H + H \rightarrow H_2$ | $7.8 \times 10^9$ |
| $H + OH \rightarrow H_2O$ | $7.0 \times 10^9$ |
| $H + H_2O_2 \rightarrow OH + H_2O$ | $9.0 \times 10^7$ |
| $H + O_2 \rightarrow HO_2$ | $2.1 \times 10^{10}$ |
| $H + HO_2 \rightarrow H_2O_2$ | $1.8 \times 10^{10}$ |
| $H + O_2^- \rightarrow HO_2^-$ | $1.8 \times 10^{10}$ |
| $H + O_3 \rightarrow HO_3$ | $3.8 \times 10^{10}$ |
| $OH + OH \rightarrow H_2O_2$ | $3.6 \times 10^9$ |
| $OH + HO_2 \rightarrow H_2O + O_2$ | $6.0 \times 10^9$ |
| $OH + O_2^- \rightarrow OH^- + O_2$ | $8.2 \times 10^9$ |
| $OH + H_2 \rightarrow H + H_2O$ | $4.3 \times 10^7$ |
| $OH + H_2O_2 \rightarrow HO_2 + H_2O$ | $2.7 \times 10^7$ |
| $OH + O^- \rightarrow HO_2^-$ | $2.5 \times 10^{10}$ |
| $OH + HO_2^- \rightarrow HO_2 + OH^-$ | $7.5 \times 10^9$ |
| $OH + O_3^- \rightarrow O_3 + OH^-$ | $2.6 \times 10^9$ |
| $OH + O_3^- \rightarrow O_2^- + O_2^- + H^+$ | $6.0 \times 10^9$ |
| $OH + O_3 \rightarrow HO_2 + O_2$ | $1.1 \times 10^8$ |
| $HO_2 + O_2^- \rightarrow HO_2^- + O_2$ | $8.0 \times 10^7$ |
| $HO_2 + HO_2 \rightarrow H_2O_2 + O_2$ | $7.0 \times 10^5$ |
| $HO_2 + O^- \rightarrow O_2 + OH^-$ | $6.0 \times 10^9$ |
| $HO_2 + H_2O_2 \rightarrow OH + O_2 + H_2O$ | $5.0 \times 10^{-1}$ |
| $HO_2 + HO_2^- \rightarrow OH + O_2 + OH^-$ | $5.0 \times 10^{-1}$ |
| $HO_2 + O_3^- \rightarrow O_2 + O_2 + OH^-$ | $6.0 \times 10^9$ |
| $HO_2 + O_3 \rightarrow HO_3 + O_2$ | $5.0 \times 10^8$ |
| $O_2^- + O_2^- + 2H_2O \rightarrow H_2O_2 + O_2 + 2OH^-$ | $1.0 \times 10^2$ / [$H_2O$]$^2$ $M^{-3}$ $s^{-1}$ |
| $O_2^- + O^- + H_2O \rightarrow O_2 + 2OH^-$ | $6.0 \times 10^8$ / [$H_2O$] $M^{-2}$ $s^{-1}$ |
| $O_2^- + H_2O_2 \rightarrow OH + O_2 + OH^-$ | $1.3 \times 10^{-1}$ |
| $O_2^- + HO_2^- \rightarrow O^- + O_2 + OH^-$ | $1.3 \times 10^{-1}$ |
| $O_2^- + O_3^- + H_2O \rightarrow O_2 + O_2 + 2OH^-$ | $1.0 \times 10^4$ / [$H_2O$] $M^{-2}$ $s^{-1}$ |
| $O_2^- + O_3 \rightarrow O_3^- + O_2$ | $1.5 \times 10^9$ |
| $O^- + O^- + H_2O \rightarrow HO_2^- + OH^-$ | $1.0 \times 10^9$ / [$H_2O$] $M^{-2}$ $s^{-1}$ |
| $O^- + O_2 \rightarrow O_3^-$ | $3.6 \times 10^9$ |
| $O^- + H_2 \rightarrow H + OH^-$ | $8.0 \times 10^7$ |
| $O^- + H_2O_2 \rightarrow O_2^- + H_2O$ | $5.0 \times 10^8$ |
| $O^- + HO_2^- \rightarrow O_2^- + OH^-$ | $4.0 \times 10^8$ |
| $O^- + O_3^- \rightarrow O_2^- + O_2^-$ | $7.0 \times 10^8$ |
| $O^- + O_3 \rightarrow O_2^- + O_2$ | $5.0 \times 10^9$ |
| $O_3^- \rightarrow O_2 + O^-$ | $3.3 \times 10^3$ $s^{-1}$ |
| $O_3^- + H^+ \rightarrow O_2 + OH$ | $9.0 \times 10^{10}$ |
| $HO_3 \rightarrow O_2 + OH$ | $1.1 \times 10^5$ $s^{-1}$ |

**Table S4.** Reaction kinetics for liquid $H_2O$.

Table S5 gives the kinetic model and the reaction constants for vapor phase $H_2O$ excerpted from Ibraguimova *et al.*[16]

| Chemical reaction | Rate constant [$M^{-1}\cdot s^{-1}$] |
|---|---|
| $H + H \rightarrow H_2$ | $3.36 \times 10^{10}$ $M^{-2}\cdot s^{-1}$ |
| $H + OH \rightarrow H_2O$ | $1.58 \times 10^{12}$ $M^{-2}\cdot s^{-1}$ |
| $OH + OH \rightarrow H_2O_2$ | $1.44 \times 10^{12}$ $M^{-2}\cdot s^{-1}$ |
| $O + OH \rightarrow O_2 + H$ | $2.07 \times 10^{10}$ |
| $H + O_2 \rightarrow HO_2$ | $1.64 \times 10^{10}$ $M^{-2}\cdot s^{-1}$ |
| $H + HO_2 \rightarrow H_2 + O_2$ | $3.68 \times 10^9$ |
| $H + HO_2 \rightarrow 2OH$ | $3.86 \times 10^{10}$ |
| $H + HO_2 \rightarrow H_2O + O$ | $2.40 \times 10^9$ |
| $HO_2 + HO_2 \rightarrow H_2O_2 + O_2$ | $1.80 \times 10^9$ |
| $H + O \rightarrow OH$ | $1.58 \times 10^{10}$ $M^{-2}\cdot s^{-1}$ |
| $H + H_2O_2 \rightarrow OH + H_2O$ | $2.43 \times 10^7$ |
| $OH + H_2O_2 \rightarrow HO_2 + H_2O$ | $9.00 \times 10^8$ |
| $OH + H_2 \rightarrow H + H_2O$ | $1.30 \times 10^4$ |
| $2OH \rightarrow H_2O + O$ | $8.39 \times 10^8$ |
| $2O \rightarrow O_2$ | $3.87 \times 10^8$ $M^{-2}\cdot s^{-1}$ |
| $O + H_2 \rightarrow OH + H$ | $5.11 \times 10^3$ |
| $H_2O_2 \rightarrow 2OH$ | $1.58 \times 10^{-20}$ |
| $H + H_2O \rightarrow H_2 + OH$ | $1.17 \times 10^{-4}$ |

**Table S5.** Reaction kinetics for gas $H_2O$.

Table S6 gives the kinetic model and the reaction constants for $O_2$ excerpted from Willis *et al.*[8]

| Chemical reaction | Rate constant [$M^{-1}\cdot s^{-1}$] |
|---|---|
| $O_2^+ + O_2 \rightarrow O_4^+$ | $4.2 \times 10^{10}$ |
| $O^+ + O_2 \rightarrow O_2^+ + O$ | $1.2 \times 10^{10}$ |
| $e^- + 2O_2 \rightarrow O_2^- + O_2$ | $2.0 \times 10^{10}$ |
| $O_2^- + O_2 \rightarrow O_4^-$ | $4.2 \times 10^{10}$ |
| $O^+ + e^- \rightarrow O$ | $2.0 \times 10^{13}$ |
| $O^+ + O_2^- \rightarrow O + O_2$ | $1.2 \times 10^{15}$ |
| $O^+ + O_4^- \rightarrow O + 2O_2$ | $1.2 \times 10^{15}$ |
| $O_2^+ + e- \rightarrow 2O$ | $1.0 \times 10^{14}$ |
| $O_2^+ + O_2^- \rightarrow 2O + O_2$ | $1.2 \times 10^{15}$ |
| $O_2^+ + O_4^- \rightarrow 2O + 2O_2$ | $1.2 \times 10^{15}$ |
| $O_4^+ + e- \rightarrow 2O + O_2$ | $1.0 \times 10^{14}$ |
| $O_4^+ + O_2^- \rightarrow 2O + 2O_2$ | $1.2 \times 10^{15}$ |
| $O_4^+ + O_4^- \rightarrow 2O + 3O_2$ | $1.2 \times 10^{15}$ |
| $O_2^- + O_3 \rightarrow O_3^- + O_2$ | $1.8 \times 10^{11}$ |
| $O_3^- + O_4^+ \rightarrow O_3 + 2O_2$ | $1.2 \times 10^{15}$ |
| $O + 2O_2 \rightarrow O_3 + O_2$ | $8.0 \times 10^{6}$ |
| $O + O \rightarrow O_2$ | $4.0 \times 10^{7}$ |
| $O + O_3 \rightarrow 2O_2$ | $6.0 \times 10^{6}$ |

**Table S6.** Reaction kinetics for $O_2$.

Table S7 gives the kinetic model and reaction constants for $CO_2$ excerpted from Kummler *et al.*[9]

| Chemical reaction | Rate constant [$M^{-1} \cdot s^{-1}$] |
|---|---|
| $CO^+ + CO \rightarrow C_2O_2^+$ | $8.91 \times 10^{-8}$ |
| $CO^+ + CO_2 \rightarrow C_2O_3^+$ | $9.03 \times 10^{-8}$ |
| $CO^+ + O_2 \rightarrow CO_3^+$ | $9.03 \times 10^{-8}$ |
| $CO_2^+ + CO_2 \rightarrow C_2O_4^+$ | $1.32 \times 10^{-7}$ |
| $CO_2^+ + CO \rightarrow C_2O_3^+$ | $1.81 \times 10^{-7}$ |
| $O_2^+ + CO_2 \rightarrow CO_4^+$ | $1.69 \times 10^{-7}$ |
| $O_2^+ + CO \rightarrow CO_3^+$ | $1.81 \times 10^{-7}$ |
| $CO_2^+ + O_2 \rightarrow O_2^+ + CO_2$ | $6.02 \times 10^{10}$ |
| $CO^+ + CO_2 \rightarrow CO_2^+ + CO$ | $8.43 \times 10^{11}$ |
| $O^+ + O_2 \rightarrow O_2^+ + O$ | $1.21 \times 10^{10}$ |
| $CO^+ + O_2 \rightarrow O_2^+ + CO$ | $1.20 \times 10^{11}$ |
| $CO_4^+ + CO \rightarrow CO_3^+ + CO_2$ | $6.02 \times 10^{7}$ |
| $C_2O_4^+ + O_2 \rightarrow O_2^+ + 2CO_2$ | $9.03 \times 10^{10}$ |
| $CO_3^+ + CO \rightarrow CO_2^+ + CO_2$ | $6.02 \times 10^{11}$ |
| $C_2O_4^+ + O_2 \rightarrow O_2^+ + CO_2$ | $2.71 \times 10^{-8}$ |
| $CO_3^+ \rightarrow O_2^+ + CO$ | $1.20 \times 10^{11}$ |
| $C_2O_2^+ + O_2 \rightarrow O_2^+ + 2CO$ | $6.02 \times 10^{8}$ |
| $C_2O_2^+ \rightarrow CO^+ + CO$ | $1.26 \times 10^{9}$ |
| $C_2O_2^+ + CO_2 \rightarrow C_3O_4^+$ | $8.43 \times 10^{-11}$ |
| $O^+ + CO_2 \rightarrow O_2^+ + CO$ | $1.81 \times 10^{11}$ |
| $C^+ + CO_2 \rightarrow CO^+ + CO$ | $9.64 \times 10^{11}$ |
| $C^+ + O_2 \rightarrow CO^+ + O$ | $6.62 \times 10^{11}$ |
| $C_2O_4^+ + CO \rightarrow C_2O_3^+ + CO_2$ | $1.32 \times 10^{11}$ |
| $C_2O_3^+ + CO \rightarrow C_2O_2^+ + CO_2$ | $1.20 \times 10^{11}$ |
| $C_2O_3^+ + O_2 \rightarrow O_2^+ + CO + CO_2$ | $6.02 \times 10^{9}$ |
| $C_2O_4^+ + CO \rightarrow C_2O_3^+ + CO_2$ | $1.81 \times 10^{-7}$ |
| $C_2O_3^+ + CO \rightarrow C_2O_2^+ + CO_2$ | $1.81 \times 10^{-7}$ |
| $e^- + O_2 + CO_2 \rightarrow CO_4^-$ | $1.99 \times 10^{-9}$ |
| $C_3O_4^+ + e^- \rightarrow 2CO + CO_2$ | $1.20 \times 10^{15}$ |
| $CO^+ + e^- \rightarrow C + O$ | $3.87 \times 10^{14}$ |
| $CO_2^+ + e^- \rightarrow CO + O$ | $1.21 \times 10^{14}$ |
| $O_2^+ + e^- \rightarrow 2O$ | $1.27 \times 10^{14}$ |
| $CO_3^+ + e^- \rightarrow CO + O_2$ | $6.08 \times 10^{14}$ |
| $CO_3^+ + e^- \rightarrow CO_2 + O$ | $6.02 \times 10^{14}$ |
| $CO_4^+ + e^- \rightarrow CO_2 + O_2$ | $6.08 \times 10^{12}$ |
| $CO_4^+ + e^- \rightarrow CO_2 + 2O$ | $6.08 \times 10^{14}$ |
| $CO_4^+ + e^- \rightarrow CO + O + O_2$ | $6.08 \times 10^{12}$ |
| $C_2O_2^+ + e^- \rightarrow 2CO$ | $3.03 \times 10^{14}$ |
| $C_2O_4^+ + e^- \rightarrow CO + O + CO_2$ | $1.21 \times 10^{15}$ |
| $C_2O_3^+ + e^- \rightarrow 2CO + O$ | $1.21 \times 10^{15}$ |
| $CO_2^+ + e^- \rightarrow CO + O$ | $1.86 \times 10^{2}$ |
| $CO^+ + e^- \rightarrow C + O$ | $1.24 \times 10^{1}$ |
| $O^+ + e^- \rightarrow O$ | $6.20 \times 10^{0}$ |
| $O_2^+ + e^- \rightarrow 2O$ | $1.86 \times 10^{2}$ |
| $C^+ + e^- \rightarrow C$ | $6.20 \times 10^{0}$ |
| $CO_3^+ + e^- \rightarrow CO + O_2$ | $1.86 \times 10^{2}$ |
| $CO_3^+ + e^- \rightarrow CO_2 + O$ | $1.86 \times 10^{0}$ |
| $CO_4^+ + e^- \rightarrow CO_2 + O_2$ | $1.86 \times 10^{0}$ |
| $CO_4^+ + e^- \rightarrow CO + O_2$ | $1.75 \times 10^{-1}$ |
| $CO_4^+ + e^- \rightarrow 2O$ | $1.75 \times 10^{2}$ |
| $C_2O_2^+ + e^- \rightarrow 2CO$ | $1.81 \times 10^{2}$ |
| $C_3O_4^+ + e^- \rightarrow 2CO$ | $1.81 \times 10^{2}$ |
| $C_2O_3^+ + e^- \rightarrow CO + CO_2$ | $1.81 \times 10^{-9}$ |
| $C_2O_3^+ + e^- \rightarrow 2CO$ | $1.81 \times 10^{2}$ |
| $C_2O_4^+ + e^- \rightarrow CO + O$ | $1.75 \times 10^{2}$ |
| $C_3O_4^+ + CO_4^- \rightarrow 2CO + CO_2$ | $1.81 \times 10^{14}$ |
| $CO^+ + CO_4^- \rightarrow CO + O_2 + CO_2$ | $1.81 \times 10^{14}$ |
| $CO_2^+ + CO_4^- \rightarrow 2CO_2 + O_2$ | $1.81 \times 10^{14}$ |
| $O_2^+ + CO_4^- \rightarrow O_2 + CO + O$ | $1.81 \times 10^{12}$ |
| $O_2^+ + CO_4^- \rightarrow 2O_2 + CO_2$ | $1.81 \times 10^{14}$ |
| $CO_3^+ + CO_4^- \rightarrow CO + 2O_2$ | $1.81 \times 10^{12}$ |
| $CO_3^+ + CO_4^- \rightarrow CO_2 + O + O_2$ | $1.81 \times 10^{14}$ |
| $C_2O_4^+ + CO_4^- \rightarrow 2CO_2 + O_2$ | $1.81 \times 10^{14}$ |
| $C_2O_2^+ + CO_4^- \rightarrow 2CO + O_2$ | $6.02 \times 10^{13}$ |
| $C_2O_3^+ + CO_4^- \rightarrow CO + CO_2 + O_2$ | $6.02 \times 10^{13}$ |
| $CO_4^+ + CO_4^- \rightarrow CO_2 + 2O_2$ | $6.02 \times 10^{13}$ |
| $C + CO_2 \rightarrow 2CO$ | $6.02 \times 10^{1}$ |
| $2O \rightarrow O_2$ | $1.77 \times 10^{-12}$ |
| $C + O_2 \rightarrow CO + O$ | $1.81 \times 10^{10}$ |
| $C_2O_3^+ + O_2 \rightarrow O_2^+ + CO$ | $1.81 \times 10^{-9}$ |
| $C_2O_2^+ + O_2 + CO_2 \rightarrow O_2^+ + 2CO$ | $1.81 \times 10^{-9}$ |
| $CO_3^+ + CO \rightarrow CO_2^+ + CO_2$ | $1.81 \times 10^{-8}$ |
| $CO_4^+ + CO \rightarrow CO_3^+ + CO_2$ | $1.81 \times 10^{-8}$ |
| $C_2O_2^+ + e^- \rightarrow C + CO_2$ | $1.81 \times 10^{-9}$ |
| $C_2O_2^+ + e^- \rightarrow C + O + CO$ | $3.01 \times 10^{12}$ |

**Table S7** Reaction kinetics for $CO_2$.

Table S8 gives the kinetic model and reaction constants in air excerpted from Willis *et al.*[17]

| Chemical reaction | Rate constant [$M^{-1}\cdot s^{-1}$] |
|---|---|
| $N_2^+ + O_2 \rightarrow N_2 + O_2^+$ | $2.95 \times 10^{11}$ |
| $O_2^+ + N_2 \rightarrow NO^+ + NO$ | $6.02 \times 10^{5}$ |
| $N_2^+ + 2N_2 \rightarrow N_4^+ + N_2$ | $6.02 \times 10^{-8}$ |
| $N_4^+ + O_2 \rightarrow O_2^+ + 2N_2$ | $6.02 \times 10^{10}$ |
| $N_2^+ + NO \rightarrow NO^+ + N_2$ | $3.01 \times 10^{11}$ |
| $O_2^+ + NO \rightarrow NO^+ + O_2$ | $4.82 \times 10^{11}$ |
| $NO^+ + e^- \rightarrow N + O$ | $2.41 \times 10^{14}$ |
| $O_2^+ + e^- \rightarrow 2O$ | $1.20 \times 10^{14}$ |
| $O_2^+ + O_2^- \rightarrow 2O + O_2$ | $1.20 \times 10^{15}$ |
| $N_2^+ + e^- \rightarrow 2N$ | $6.02 \times 10^{13}$ |
| $N_2^+ + O_2^- \rightarrow 2N + O_2$ | $1.20 \times 10^{15}$ |
| $e^- + O_2 \rightarrow O_2^-$ | $1.14 \times 10^{-9}$ |
| $e^- + O_2 \rightarrow O_2^-$ | $6.02 \times 10^{-11}$ |
| $e^- + NO_2 \rightarrow NO_2^-$ | $2.41 \times 10^{10}$ |
| $e^- + NO \rightarrow NO^-$ | $7.83 \times 10^{-11}$ |
| $O_2^- + O_3 \rightarrow O_3^- + O_2$ | $1.81 \times 10^{11}$ |
| $O_3^- + NO \rightarrow NO_2^- + O_2$ | $4.82 \times 10^{11}$ |
| $O_3^- + NO_2 \rightarrow NO_2^- + O_3$ | $4.22 \times 10^{11}$ |
| $NO_2^- + O_3 \rightarrow NO_3^- + O_2$ | $6.02 \times 10^{9}$ |
| $NO^- + O_2 \rightarrow O_2^- + NO$ | $5.42 \times 10^{11}$ |
| $N + O \rightarrow NO$ | $5.42 \times 10^{-12}$ |
| $2N \rightarrow N_2$ | $1.26 \times 10^{-12}$ |
| $N + O_2 \rightarrow NO + O$ | $4.82 \times 10^{8}$ |
| $N + O_3 \rightarrow NO + O_2$ | $3.43 \times 10^{8}$ |
| $N + NO \rightarrow N_2 + O$ | $1.32 \times 10^{10}$ |
| $N + NO_2 \rightarrow N_2O + O$ | $4.64 \times 10^{9}$ |
| $N + NO_2 \rightarrow 2NO$ | $3.55 \times 10^{9}$ |
| $N + NO_2 \rightarrow N_2 + O_2$ | $1.08 \times 10^{9}$ |
| $N + NO_2 \rightarrow N_2 + 2O$ | $1.39 \times 10^{9}$ |
| $O + O_2 \rightarrow O_3$ | $1.69 \times 10^{-13}$ |
| $O + O \rightarrow O_2$ | $1.20 \times 10^{-12}$ |
| $O + O_3 \rightarrow 2O_2$ | $6.02 \times 10^{6}$ |
| $O + NO \rightarrow NO_2$ | $4.88 \times 10^{-11}$ |
| $O + NO \rightarrow NO_2$ | $6.62 \times 10^{-11}$ |
| $O + NO_2 \rightarrow NO + O_2$ | $1.51 \times 10^{10}$ |
| $O_2 + 2NO \rightarrow 2NO_2$ | $6.02 \times 10^{-13}$ |
| $O_3 + NO \rightarrow NO_2 + O_2$ | $2.11 \times 10^{8}$ |

**Table S8.** Reaction kinetics for air.

Supplementary Text IV: MATLAB script explanation

The MATLAB scripts and custom functions are available on GitHub.[4] To solve the Partial Differential Equations (PDEs) shown as Equation 5 in the main text, MATLAB's "PDEPE" solver was used,[18] which solves 1-D parabolic and elliptic PDEs using the ode15s method and is optimized for stiff ordinary differential equations (ODEs) with a low to medium accuracy level. The "PDEPE" solver has specific input for $c(x, t, u, \partial u/\partial x)$, $f(x, t, u, \partial u/\partial x)$, and $s(x, t, u, \partial u/\partial x)$, which are defined as

$$c\left(x, t, u, \frac{\partial u}{\partial x}\right)\frac{\partial u}{\partial t} = x^{-m}\frac{\partial}{\partial x}\left(x^m f\left(x, t, u, \frac{\partial u}{\partial x}\right)\right) + s\left(x, t, u, \frac{\partial u}{\partial x}\right) \qquad (S4)$$

Equation 5 ($\mathrm{v} = 0$) in the main text is then converted into the form Equation S4:

$$\frac{1}{D_i}\cdot\frac{\partial C_i}{\partial t} = x^0\frac{\partial}{\partial x}\left(x^0\frac{\partial C_i}{\partial x}\right) + \frac{1}{D_i}\left(R_i - \sum_j k_{ij}C_iC_j + \sum_{j,k\neq i} k_{jk}C_jC_k\right) \qquad (S5)$$

This leads to input parameters as follows

$$c\left(x, t, u, \frac{\partial u}{\partial x}\right) = \frac{1}{D_i} \qquad (S6)$$

$$f\left(x, t, u, \frac{\partial u}{\partial x}\right) = \frac{\partial C_i}{\partial x} \qquad (S7)$$

and

$$s\left(x, t, u, \frac{\partial u}{\partial x}\right) = \frac{1}{D_i}\left(R_i - \sum_j k_{ij}C_iC_j + \sum_{j,k\neq i} k_{jk}C_jC_k\right) \qquad (S8)$$

The parameter $m$ in Equation S4 is taken to be 0. The parameters can be implemented in the script as follows:

```
function [c,f,s] = pdefun(x,t,C,dCdx)
     c = [1/D(1);1/D(2);1/D(3); ……];
     f = [dCdx(1);dCdx(2);dCdx(3); ……];

     n = rem(floor(t/dwell_time),10);

     if x>=(45.45+n) && x<=(45.55+n)
         s = [R(1)/D(1);R(2)/D(2);R(3)/D(3); ……];
     else
         s = [0;0;0;0;0;0;0;0;0;0;0;0;0;0;0;0];
     end

     k = 1e-6 .* [2.95e11;6.02e5;6.02e-8; ……];

     r = zeros(37,1);
     r(1) = k(1) * C(5);
     r(2) = k(2) * C(10);
     r(3) = k(3) * C(5) * C(4);
      …

     products = zeros(20,1);
     products(1) = -r(7) - r(8) - r(10) - r(12) - r(13)
     - r(14) - r(15);
     products(2) = r(7) + 2 * r(10) + 2 * r(11) - r(21)
     - 2 * r(22) - r(23) - r(24) - r(25) - r(26) -
     r(27) - r(28) - r(29);
     products(3) = 0;
     …

     s(1) = s(1) + products(1)/D(1);
     s(2) = s(2) + products(2)/D(2);
     s(3) = s(3) + products(3)/D(3);
     …
end
```

**Script S1.** Pseudo-code for "PDEPE" function.

The "PDEPE" function requires initial and boundary conditions to solve the equation. The initial condition C0 = icfun($x$) is a function containing a matrix of concentration information C[$N$] at 1-D position $x$. To set C0, the matrix ic($N$, $x$ = 1001) containing the initial spatial concentration is passed as an argument and converted using the following script.

```
function C0 = icfun(x,ic)
     C0=[ic(1,round(x*10) + 1);ic(2,round(x*10) +
1);ic(3,round(x*10) + 1) …… ];
end
```

**Script S2.** Initial condition function for "PDEPE". Range of x for C0 in the code is 0 to 100.

For the boundary conditions, we assume the initial gas species have fixed concentration at the left and right boundaries of the system, and the other species have zero concentration at the left and right boundaries, regardless of time. The boundary condition $p(x,t,u)$ and $q(x,t)$ in "PDEPE" is defined in Equation S9:

$$p(x,t,u) + q(x,t)f\left(x,t,u,\frac{\partial u}{\partial x}\right) = 0 \qquad (S9)$$

For the left boundary ($x$ = 0), the boundary function for water in 10 torr gas conditions, $C_{H2O}(0,t) = 538$ μM can be written as Equation S10:

$$C_{H2O}(0,t) - 538\ [\mu\text{M}] + 0 \cdot f\left(x,t,u,\frac{\partial u}{\partial x}\right) = 0 \qquad (S10)$$

For the right boundary ($x$ = $L$), the boundary function of water in 10 torr gas conditions, $C_{H2O}(L,t) = 538\ \mu M$ can be written as Equation S11:

$$C_{H2O}(L,t) - 538\ [\mu\text{M}] + 0 \cdot f\left(x,t,u,\frac{\partial u}{\partial x}\right) = 0 \qquad (S11)$$

For other species having zero initial concentration, their left and right boundary functions are:

$$C(0,t) + 0 \cdot f\left(x,t,u,\frac{\partial u}{\partial x}\right) = 0 \qquad (S12)$$

$$C(L,t) + 0 \cdot f\left(x,t,u,\frac{\partial u}{\partial x}\right) = 0 \qquad (S13)$$

Therefore, the implemented code for the boundary function is shown in Script S3.

```
function [pL,qL,pR,qR] = bcfun(xL,CL,xR,CR,t)
     pR = CR;
     pR(16) = CR(16)-538;
     qR = [0;0;0;0;0;0;0;0;0;0;0;0;0;0;0;0];
     pL = CL;
     pL(16) = CL(16)-538;
     qL = [0;0;0;0;0;0;0;0;0;0;0;0;0;0;0;0];
end
```

**Script S3.** Boundary condition function for "PDEPE".

Supplementary Text V: Influence of advection, generation, and kinetics toward diffusion

Fluid circulation affects mass transport during radiolysis. To estimate the influence of flow in the presence of diffusion in the fluid, the mass transport term can be re-established using the advection-diffusion equation:

$$\frac{\partial C}{\partial t} = \nabla \cdot (D\nabla C - \mathrm{v}C) + R \qquad (S14)$$

In the above equation, *C* is the concentration, *D* is the diffusivity, v is the linear velocity, and *R* is the source and sink term, which are the generation and reaction terms in this case. The Equation S14 can be re-written under the radiolysis reaction condition in 1-D as follows:

$$\frac{\partial C_i}{\partial t} = R_i - \sum_j k_{ij} C_i C_j + \sum_{j,k \neq i} k_{jk} C_j C_k + \frac{\partial}{\partial x}\left(D \frac{\partial C_i}{\partial x} - \mathrm{v}C_i\right) \qquad (S15)$$

To compare the contribution of advection and diffusion toward mass transport, Equation S15 can be nondimensionalized by introducing the following variables:[19]

$$\tilde{x} = \frac{x}{l},\ \tilde{t} = \frac{t}{\Delta t},\ \tilde{C}_i = \frac{C_i}{C_{ssi}},\ \tilde{C}_j = \frac{C_j}{C_{ssj}},\ \text{and}\ \tilde{C}_k = \frac{C_k}{C_{ssk}} \qquad (S16)$$

with $l$ as the characteristic length scale, $t$ being the time scale, and $C_{ssi}$, $C_{ssj}$, $C_{ssk}$ being the steady-state concentrations of species $i$, $j$, and $k$, respectively. Subsequently, Equation S16 can be converted into the following nondimensionalized equation:

$$\frac{\partial \tilde{C}_i}{\partial \tilde{t}}\frac{l^2}{\Delta t D_i} = \frac{l^2}{D_i C_{ssi}} R_i - \frac{\sum_j k_{ij} C_{ssj} l^2}{D_i} \tilde{C}_i \tilde{C}_j + \frac{\sum_{j,k \neq i} k_{jk} C_{ssj} C_{ssk} l^2}{D_i C_{ssi}} \tilde{C}_j \tilde{C}_k + \frac{\partial^2 \tilde{C}_i}{\partial \tilde{x}^2} - \frac{\mathrm{v} l}{D_i}\frac{\partial \tilde{C}_i}{\partial \tilde{x}} \quad (S17)$$

The coefficient for each component unrelated to convection is defined as the Damköhler number ($\mathrm{Da_I}$). For example, $\mathrm{Da_I}$ related to electron dose is $l^2/D_i C_{ssi}$. The coefficient $\mathrm{v}l/D$ related to convection is defined as the Péclet number ($\mathrm{Pe}_L$). Therefore, the relative contribution of each term to diffusion increases with the magnitude of $\mathrm{Da_I}$. Diffusion effects dominate mass transport when $\mathrm{Pe}_L \to 0$, while advection dominates when $\mathrm{Pe}_L \gg 1$. When $\mathrm{Pe}_L \approx 1$, both advection and diffusion contribute significantly to mass transport. $\mathrm{Da_I}$ and $\mathrm{Pe}_L$ for typical LPTEM and GPTEM experiments (Table S9) are calculated and compared in Table S10.

| | Liquid | Gas |
|---|---|---|
| $\varphi$ [Gy/s] | $10^5$-$10^9$ | |
| $l$ [m] | $10^{-5}$ | |
| $k$ [M/s] | $10^8$-$10^{10}$ | |
| $G$ [#/100eV] | $10^0$ | |
| $\rho$ [kg/m$^3$] | $10^3$ | $10^0$ |
| $C_{ss}$ [M] | $10^{-6}$-$10^{-3}$ | $10^{-11}$-$10^{-9}$ |
| $D$ [m$^2$/s] | $10^{-9}$ | $10^{-3}$ |
| v [m/s] | $10^{-4}$-$10^{-2}$ | $10^{-2}$-$10^0$ |
| $R_i$ | $10^1$-$10^5$ | $10^{-2}$-$10^2$ |

**Table S9.** Environmental parameters for typical LPTEM and GPTEM experiments.

|  | $\mathrm{Da_I}$ (dose) | $\mathrm{Da_I}$ (kinetics) | $\mathrm{Pe}_L$ |
|---|---|---|---|
| Liquid | $10^{3}$-$10^{10}$ | $10^{1}$-$10^{6}$ | $10^{0}$-$10^{2}$ |
| Gas | $10^{2}$-$10^{4}$ | $10^{-10}$-$10^{-6}$ | $10^{-4}$-$10^{-2}$ |

**Table S10.** Calculated $\mathrm{Da_I}$ and $\mathrm{Pe}_L$ for typical LPTEM and GPTEM experiments.

In our liquid cell experiments, the volumetric flow rate of the experimental fluid ($Q$) is in the range of 1-2 μL/min at most, which yields an average linear flow rate of 0.017 m/s with a 1 μm channel gap. However, as there is flow resistance and a large bypass volume in the liquid cell holder, the velocity is lower in reality. On the other hand, the maximum achievable gas flow rate with the Protochips Atmosphere system is around 1 sccm (standard cubic centimeters per minute), which converts to a flow velocity of 0.835 m/s at a 10 μm channel gap. By substituting the length scale of the viewing area in TEM ($l \approx 10\ \mu$m), and the diffusivity of the liquid ($D_l \approx 10^{-9}\ \mathrm{m^2/s}$) and gas ($D_g \approx 10^{-3}\ \mathrm{m^2/s}$), the Péclet number in LPTEM is ≈170 while it is ≈0.08 in GPTEM. This implies that typical advection in LPTEM affects the radicals' concentration but does not impact GPTEM conditions.

The MATLAB implementation of Equation S15 can be achieved by transforming it into the MATLAB "PDEPE" solver input form shown in Equation S4:

$$\frac{1}{D}\cdot\frac{\partial C_i}{\partial t} = x^0\frac{\partial}{\partial x}\left(x^0\left(\frac{\partial C_i}{\partial x} - \frac{\mathrm{v}}{D}C_i\right)\right) + \frac{1}{D}\left(R_i - \sum\nolimits_j k_{ij}C_iC_j + \sum\nolimits_{j,k\neq i} k_{jk}C_jC_k\right) \qquad (S18)$$

This leads to input parameters

$$c\left(x,t,u,\frac{\partial u}{\partial x}\right) = \frac{1}{D} \qquad (S19)$$

$$f\left(x,t,u,\frac{\partial u}{\partial x}\right) = \frac{\partial C_i}{\partial x} - \frac{\mathrm{v}}{D}C_i \qquad (S20)$$

and

$$s\left(x,t,u,\frac{\partial u}{\partial x}\right)=\frac{1}{D}\left(R_i-\sum_j k_{ij}C_iC_j+\sum_{j,k\neq i}k_{jk}C_jC_k\right) \qquad (S21)$$

These parameters can be implemented into the script as follows:

```
function [c,f,s] = pdefun(x,t,C,dCdx)
      c = [1/D(1);1/D(2);1/D(3); ……];
      f = [dCdx(1)-v*C(1)/D(1);dCdx(2)-
      v*C(2)/D(2);dCdx(3)-v*C(3)/D(3); ……];

      n = rem(ceil(t/dwell_time),10);

      if x>=(45.45+n) && x<=(45.55+n)
          s = [R(1)/D(1);R(2)/D(2);R(3)/D(3); ……];
      else
          s = [0;0;0;0;0;0;0;0;0;0;0;0;0;0;0;0];
      end

      k = 1e-6 .* [2.95e11;6.02e5;6.02e-8; ……];

      r = zeros(37,1);
      r(1) = k(1) * C(5);
      r(2) = k(2) * C(10);
      r(3) = k(3) * C(5) * C(4);
       …

      products = zeros(20,1);
      products(1) = -r(7) - r(8) - r(10) - r(12) - r(13)
      - r(14) - r(15);
      products(2) = r(7) + 2 * r(10) + 2 * r(11) - r(21)
      - 2 * r(22) - r(23) - r(24) - r(25) - r(26) -
      r(27) - r(28) - r(29);
      products(3) = 0;
      …
      s(1) = s(1) + products(1)/D(1);
      s(2) = s(2) + products(2)/D(2);
      s(3) = s(3) + products(3)/D(3);
      …
end
```

**Script S4.** Pseudo-code for "PDEPE" function with consideration of advection. Advection variable is denoted in red color.

The initial and boundary conditions are kept the same with non-advection conditions. Therefore, Script S2, S3 can be used to solve this PDE as well. The results under the flow velocity of 0.001-10 m/s in LPTEM, 0.01-100 m/s in LPSTEM, and 10-1000 m/s in GPTEM are illustrated in Figures S5-7.

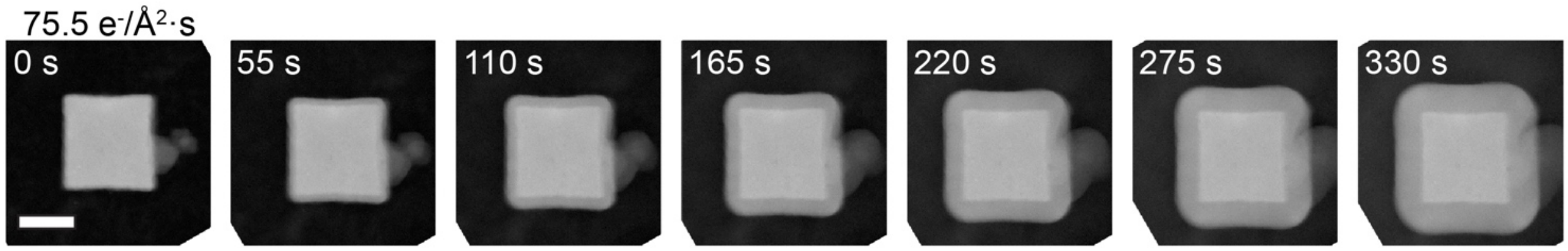

**Figure S1.** Time-series HAADF-STEM images showing the oxidation behavior of 200 nm aluminum nanocubes at 75.5 $e^{-}/Å^{2}s$ ($3.0 \times 10^{8}$ Gy/s) electron dose. 10 torr $O_2$ is introduced at room temperature. Scale bar is 100 nm.

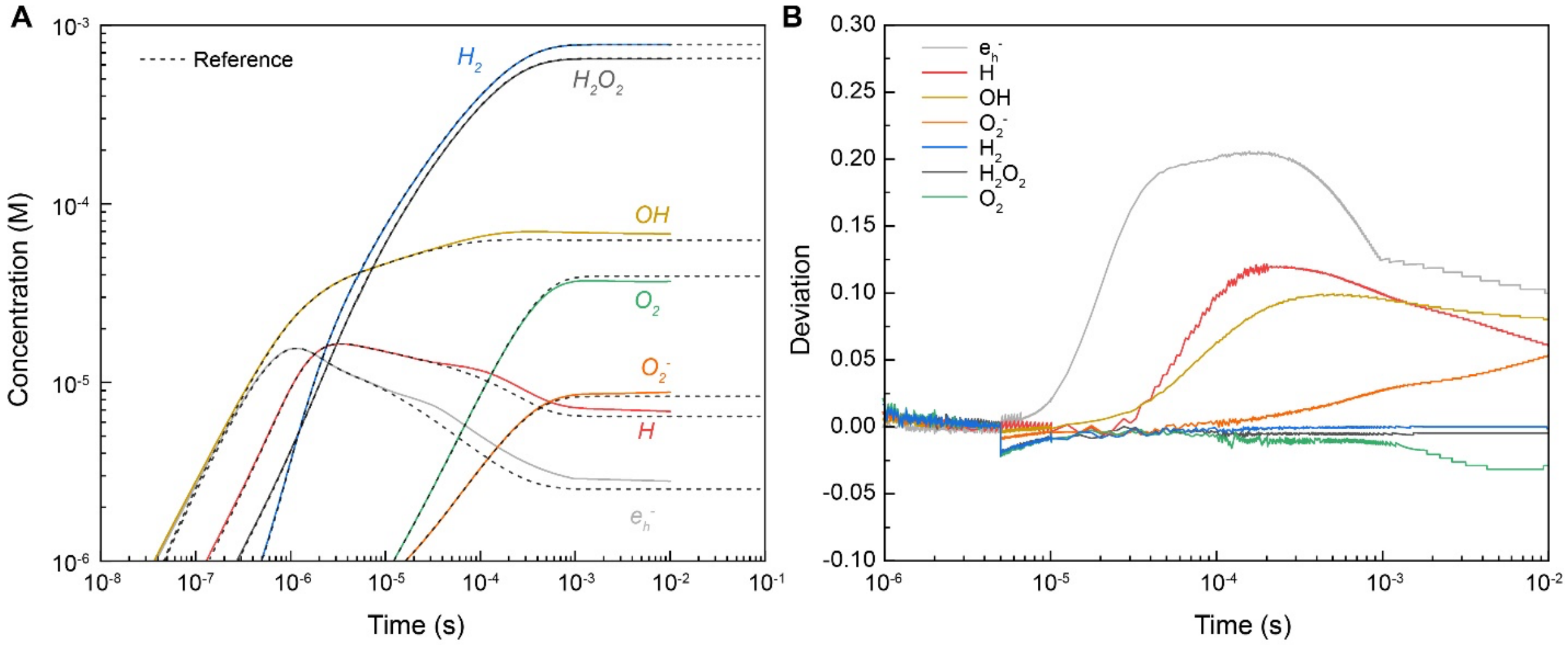

**Figure S2. Validation of our mathematical model. (A)** Radical concentration in liquid water in TEM mode with a dose rate of $7.5 \times 10^7$ Gy/s and a beam radius of 1 μm with a suppressed diffusion coefficient. The black dotted line is the result from the reference model script.[6] **(B)** The deviation between the reference and our model as a function of time.

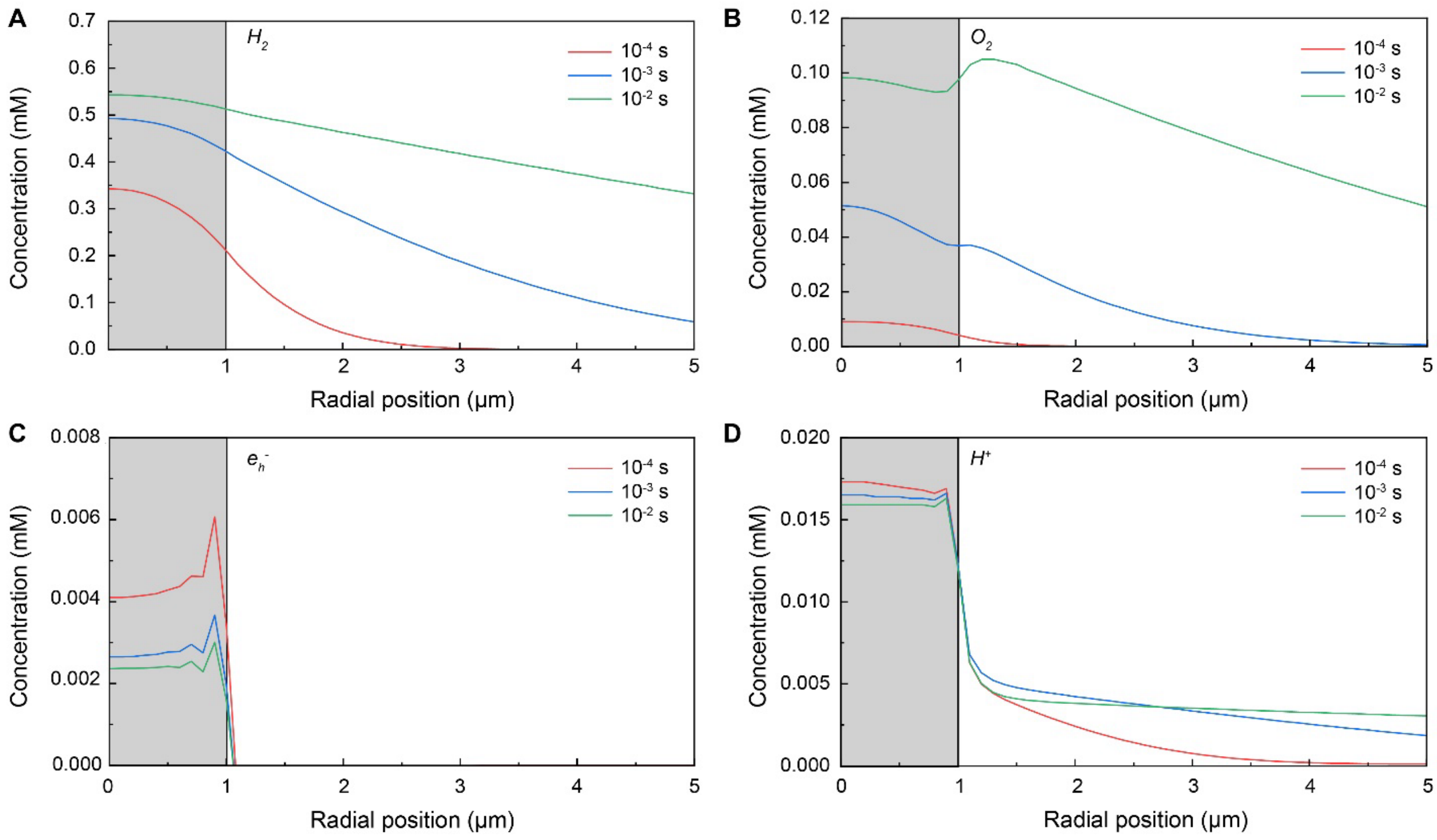

**Figure S3. Spatiotemporal distribution of radiolysis species in water. (A)** Distribution of $H_2$. **(B)** Distribution of $O_2$. **(C)** Distribution of $e_h^-$. **(D)** Distribution of $H^+$. (A-D) The simulation parameter is liquid water in TEM mode with a dose rate of $7.5 \times 10^7$ Gy/s and a beam radius of 1 μm.

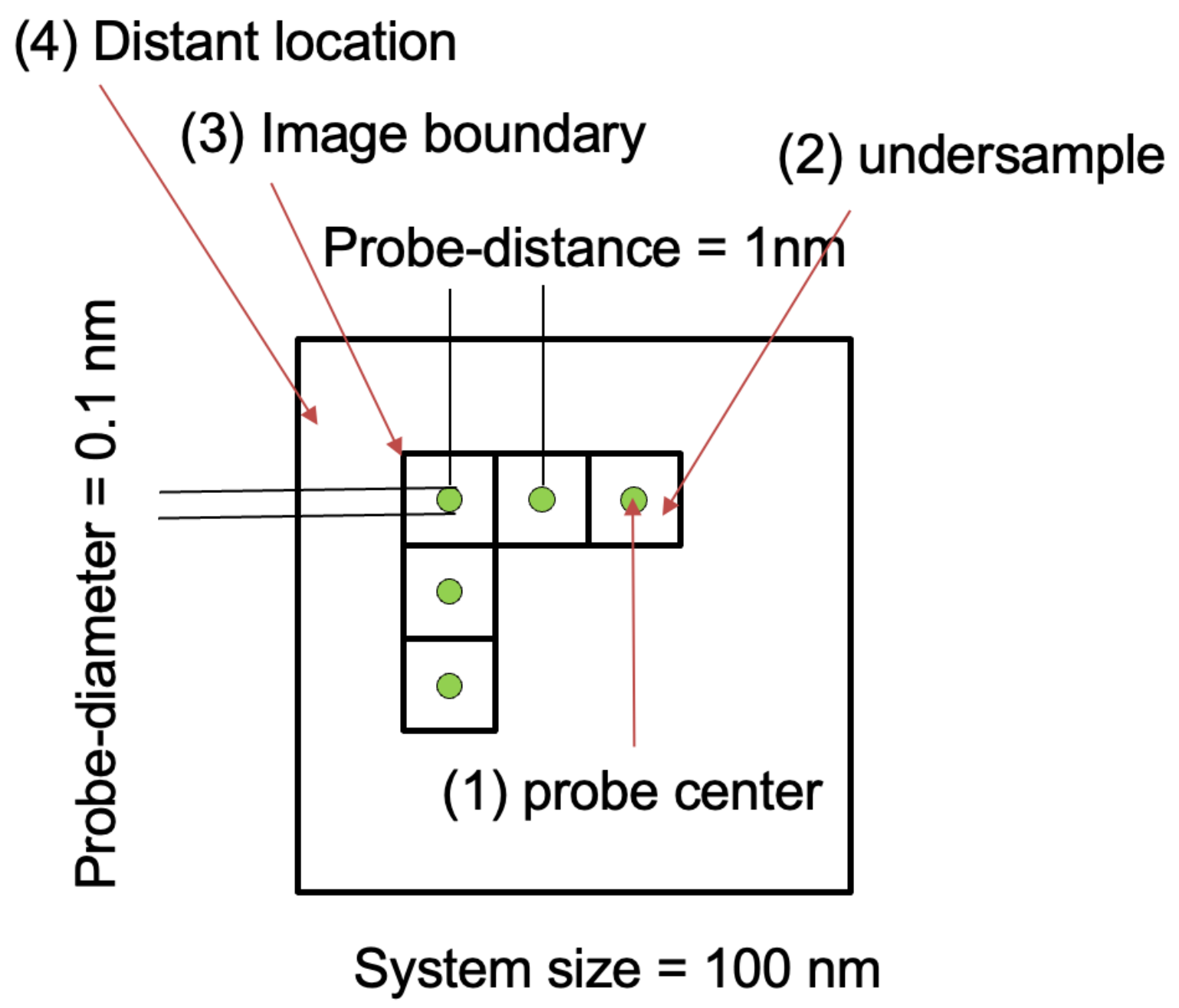

**Figure S4. Schematic definition of the calculation mesh and concentration-sampling points.**

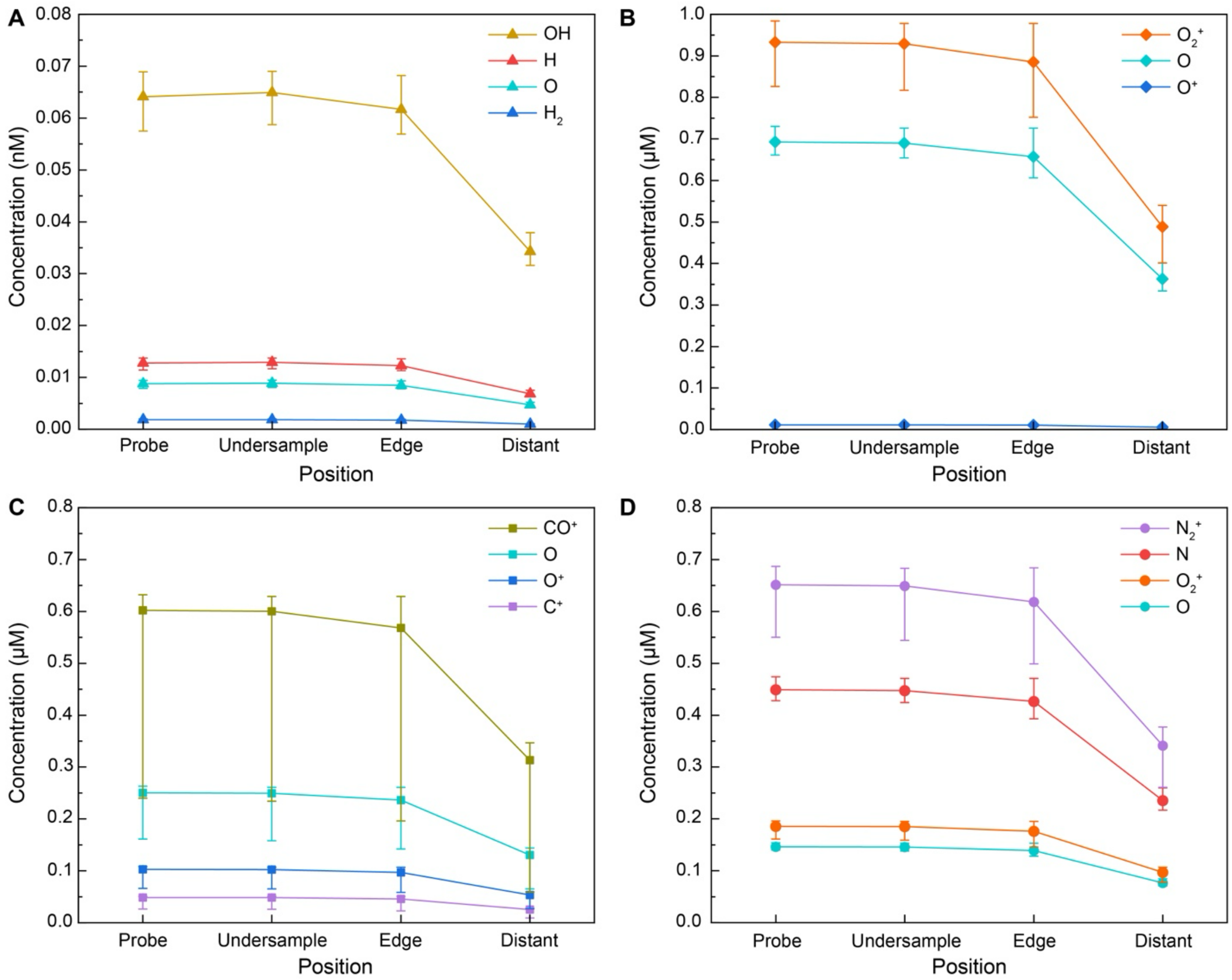

**Figure S5. Calculated concentrations of major radiolysis species in gas-phase STEM at different locations including the center of the electron probe, under-sampled area, image edge, and distant location (20 nm from the image edge)** with **(A)** 10 torr water vapor, **(B)** 760 torr $O_2$, **(C)** 760 torr $CO_2$, and **(D)** 760 torr dry air. (A-D) The probe diameter and current is 1 Å, and 62 pA respectively, yielding a dose rate of $5.8 \times 10^7$ Gy/s.

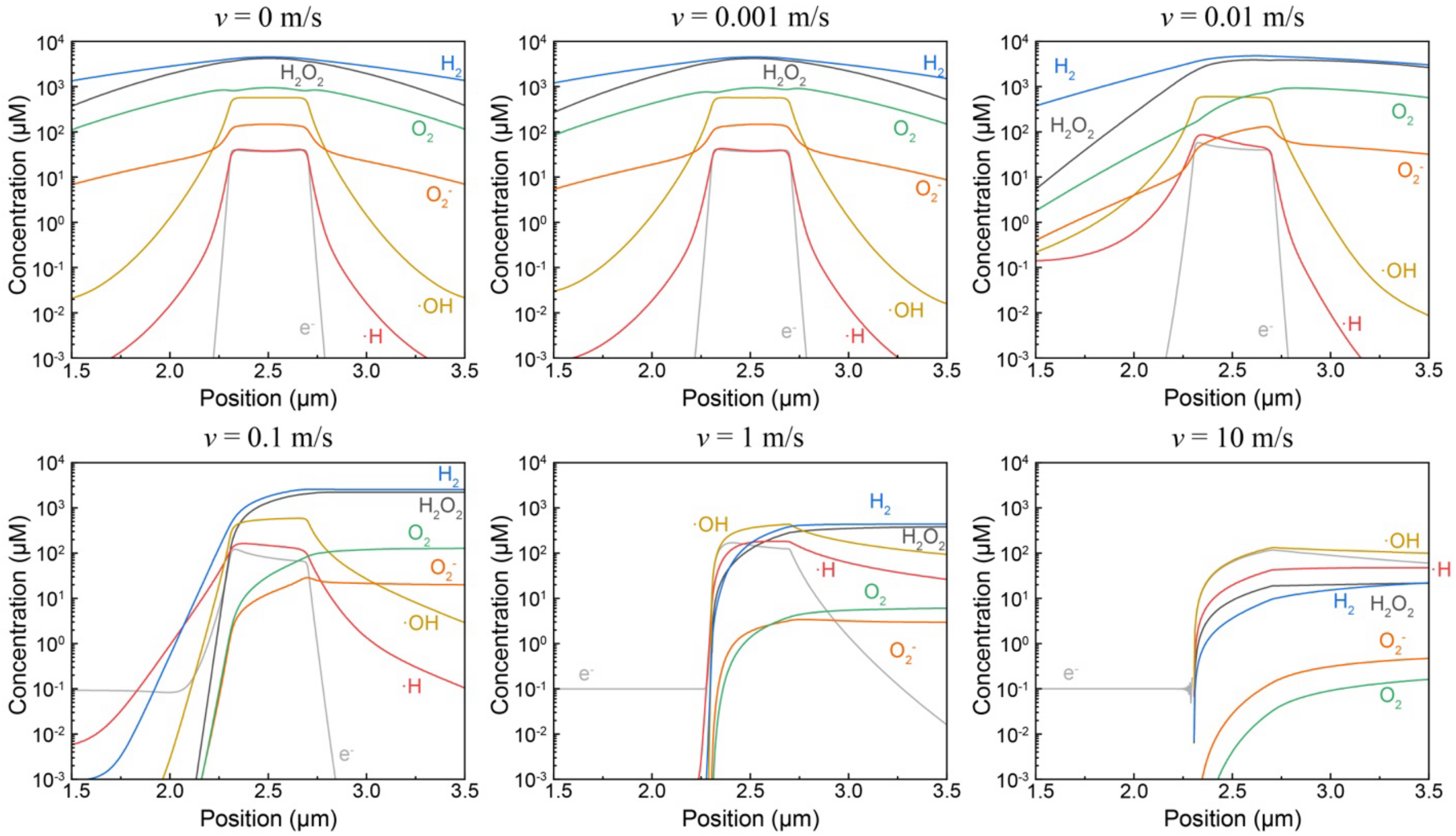

**Figure S6. Spatial distribution of radiolysis species in the liquid phase TEM under different flow conditions.** The radius of illumination is 200 nm and the corresponding dose rate is $9.4 \times 10^9$ Gy/s. *v* denotes the fluid's linear velocity.

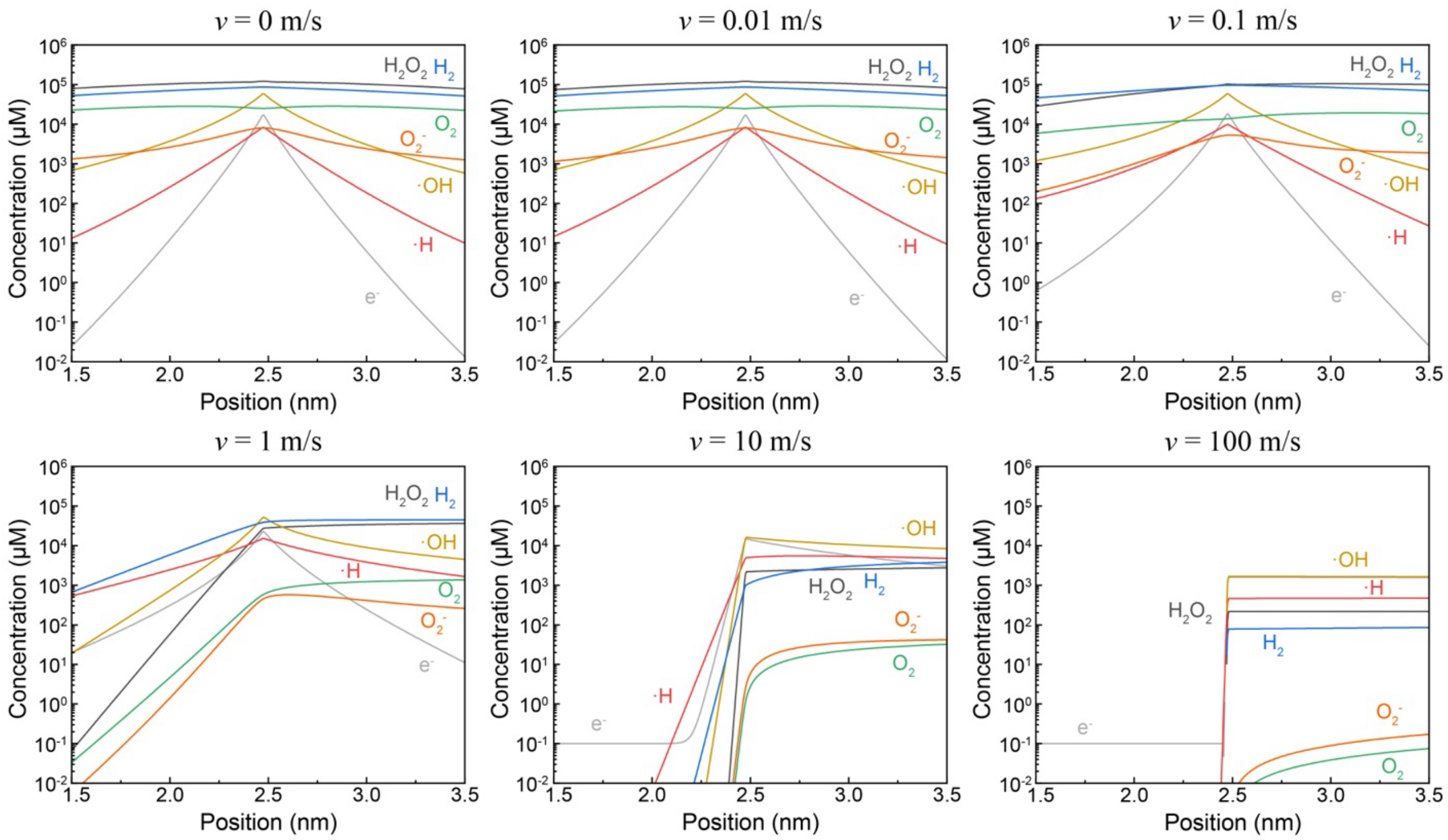

**Figure S7. Spatial distribution of radiolysis species in the liquid phase STEM under different flow conditions.** The probe diameter is 1 Å, and the probe current is 62 pA, yielding a dose rate of $5.85 \times 10^7$ Gy/s. $v$ denotes the fluid's linear velocity.

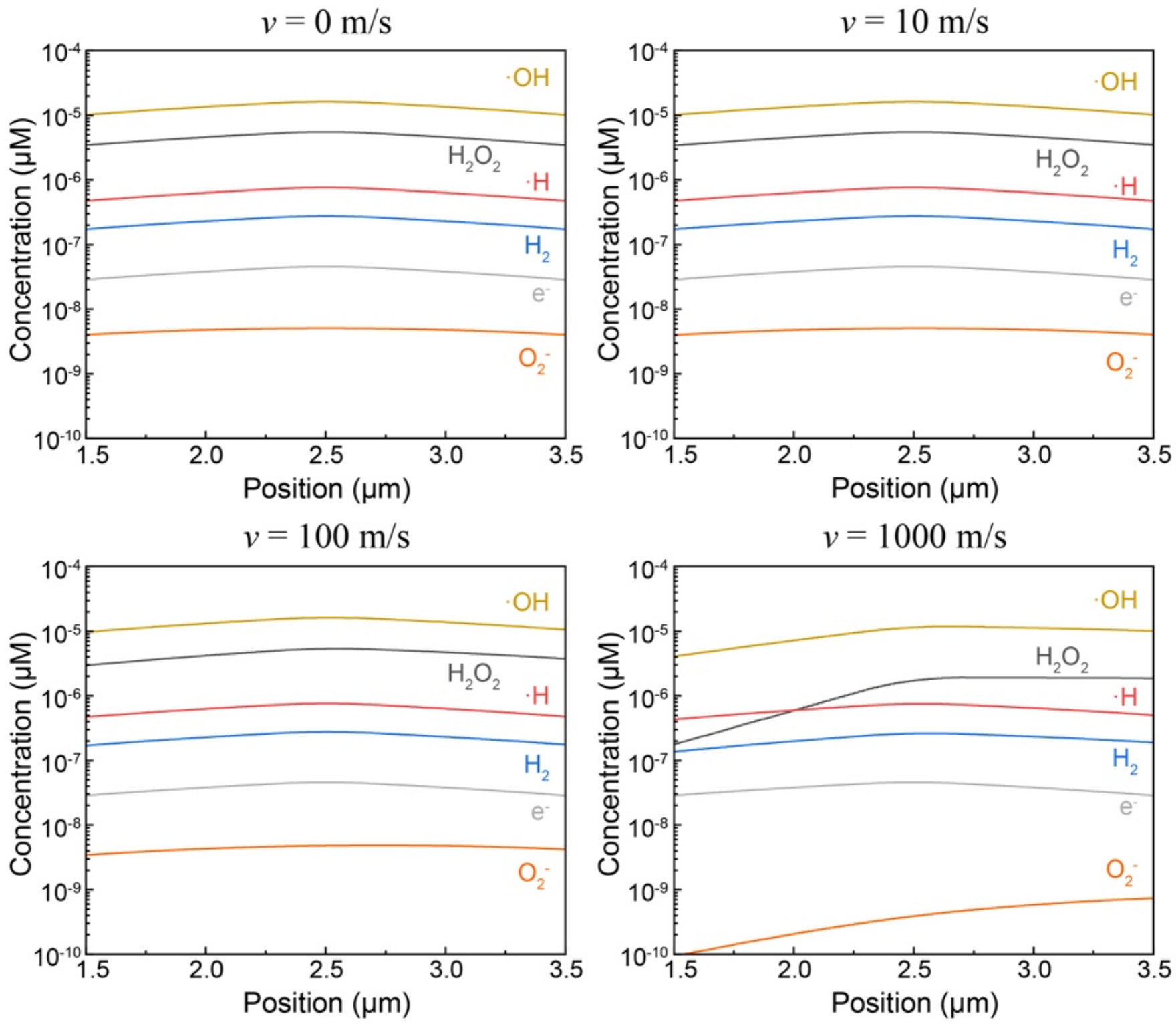

**Figure S8. Spatial distribution of radiolysis species in the gas phase TEM (10 torr, water vapor) under different flow conditions.** The radius of illumination is 200 nm and the corresponding dose rate is $9.4 \times 10^9$ Gy/s. *v* denotes the fluid's linear velocity.

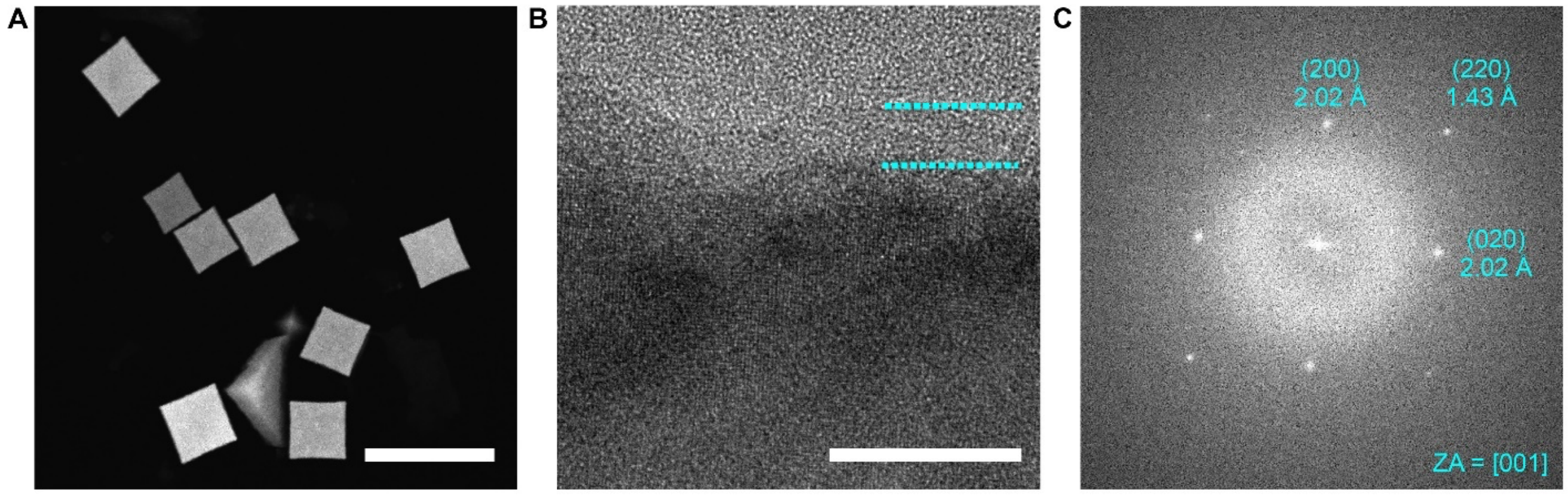

**Figure S9. Pristine aluminum nanocubes. (A)** Low-magnification high angle annular dark field (HAADF) STEM image of Al nanocubes. The scale bar is 500 nm. **(B)** High-resolution TEM image of the surface of an Al nanocube. The cyan dotted lines indicate the surface of the native oxide layer and the interface between the oxide layer and the aluminum. The native oxide layer before *in situ* oxidation is ≈ 3 nm. The scale bar is 10 nm. **(C)** Fast Fourier Transform (FFT) of image (B) shows the FCC structure.

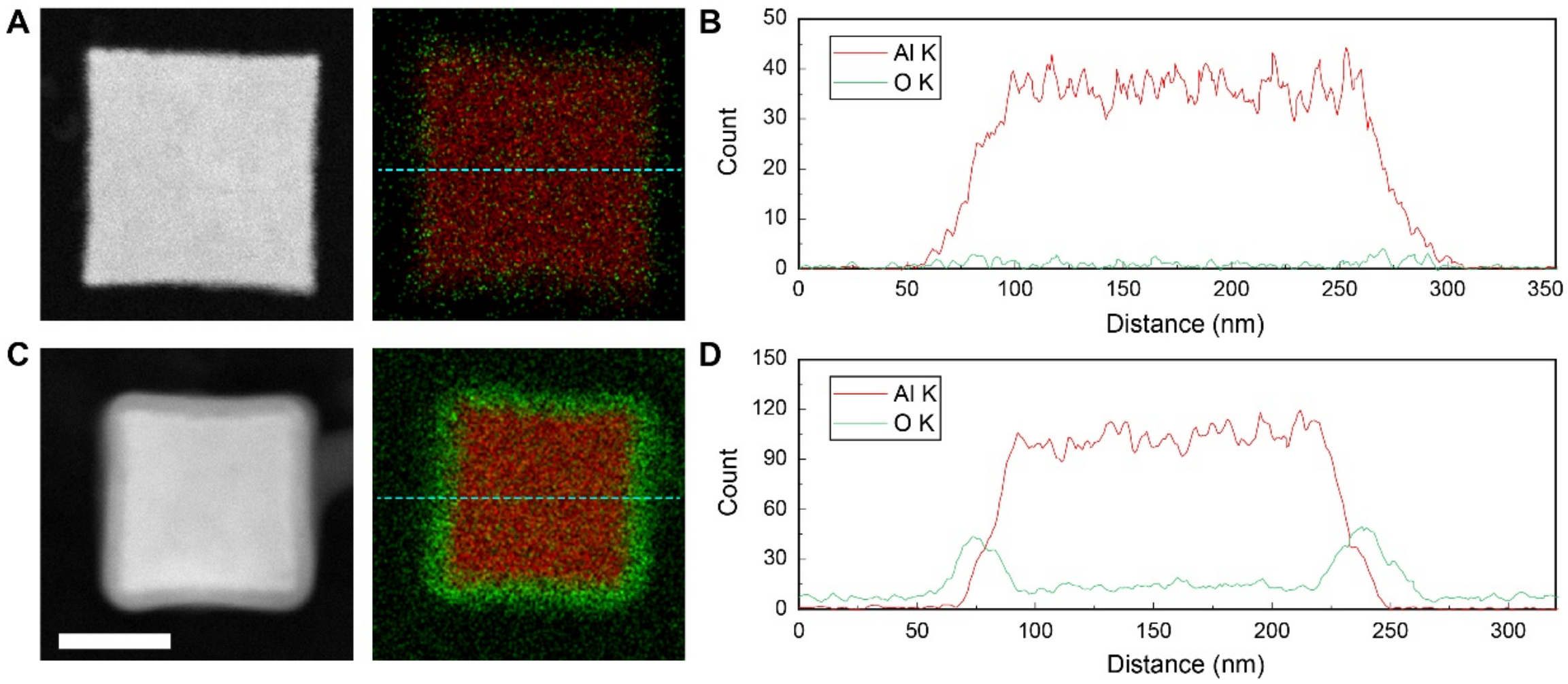

**Figure S10. Elemental maps of pristine and oxidized aluminum nanocube. (A)** EDS elemental map of a pristine nanocube. **(B)** EDS line profile of (A). **(C)** EDS elemental map of an $O_2$-oxidized Al nanocube. **(D)** EDS line profile of (C). The scale bar is 100 nm. The red and green colors indicate the signal from Al *K* and O *K*, respectively.

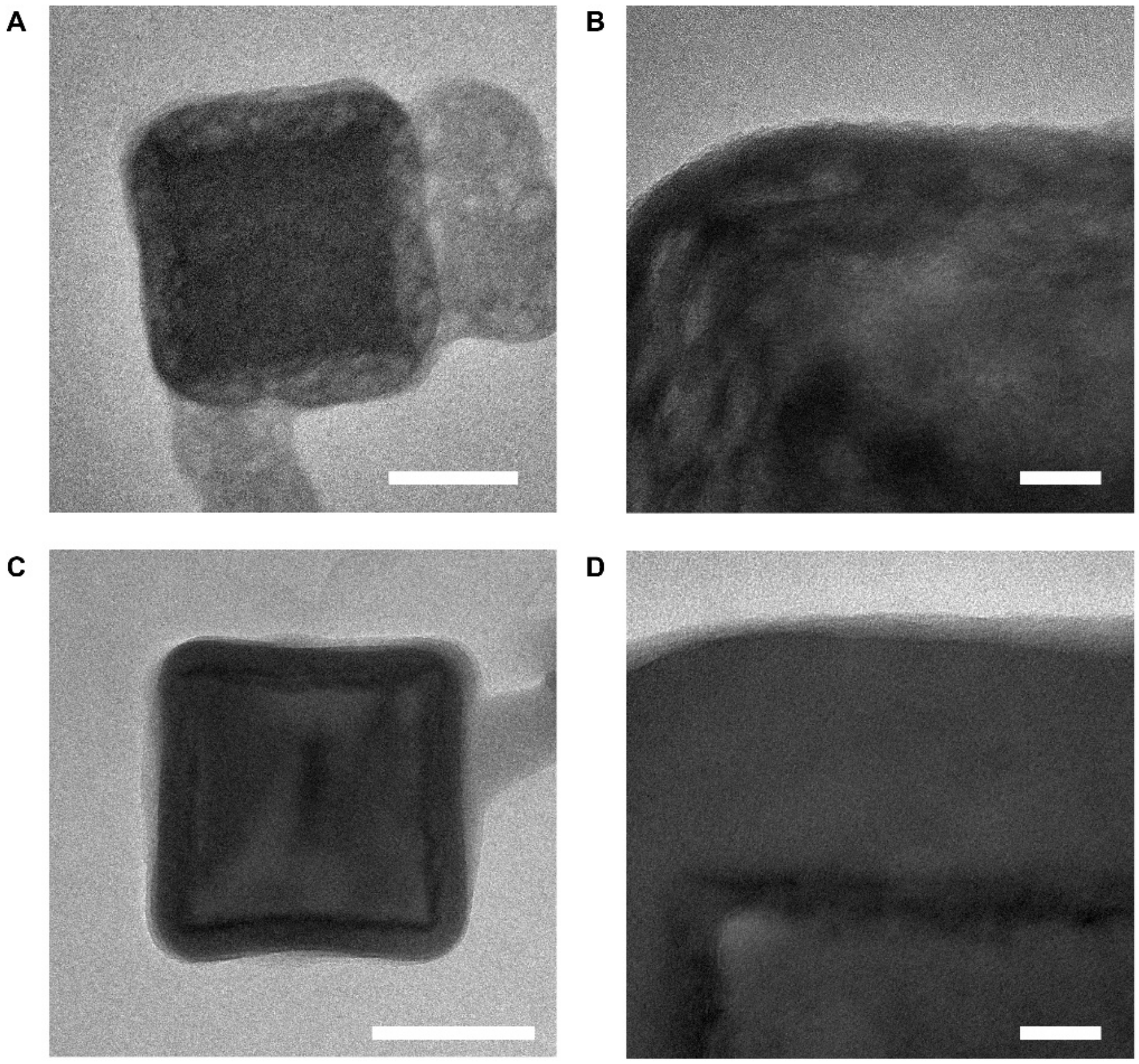

**Figure S11. Comparisons between a $H_2O$-oxidized aluminum nanocube and an $O_2$-oxidized aluminum nanocube. (A)** Low-magnification and **(B)** high-magnification TEM images of a $H_2O$-oxidized nanocube. The scale bars in (A) and (B) are 100 nm and 20 nm, respectively. **(C)** Low-magnification and **(D)** high-magnification TEM image of an $O_2$-oxidized nanocube. The scale bars in (C) and (D) are 100 nm and 20 nm, respectively.

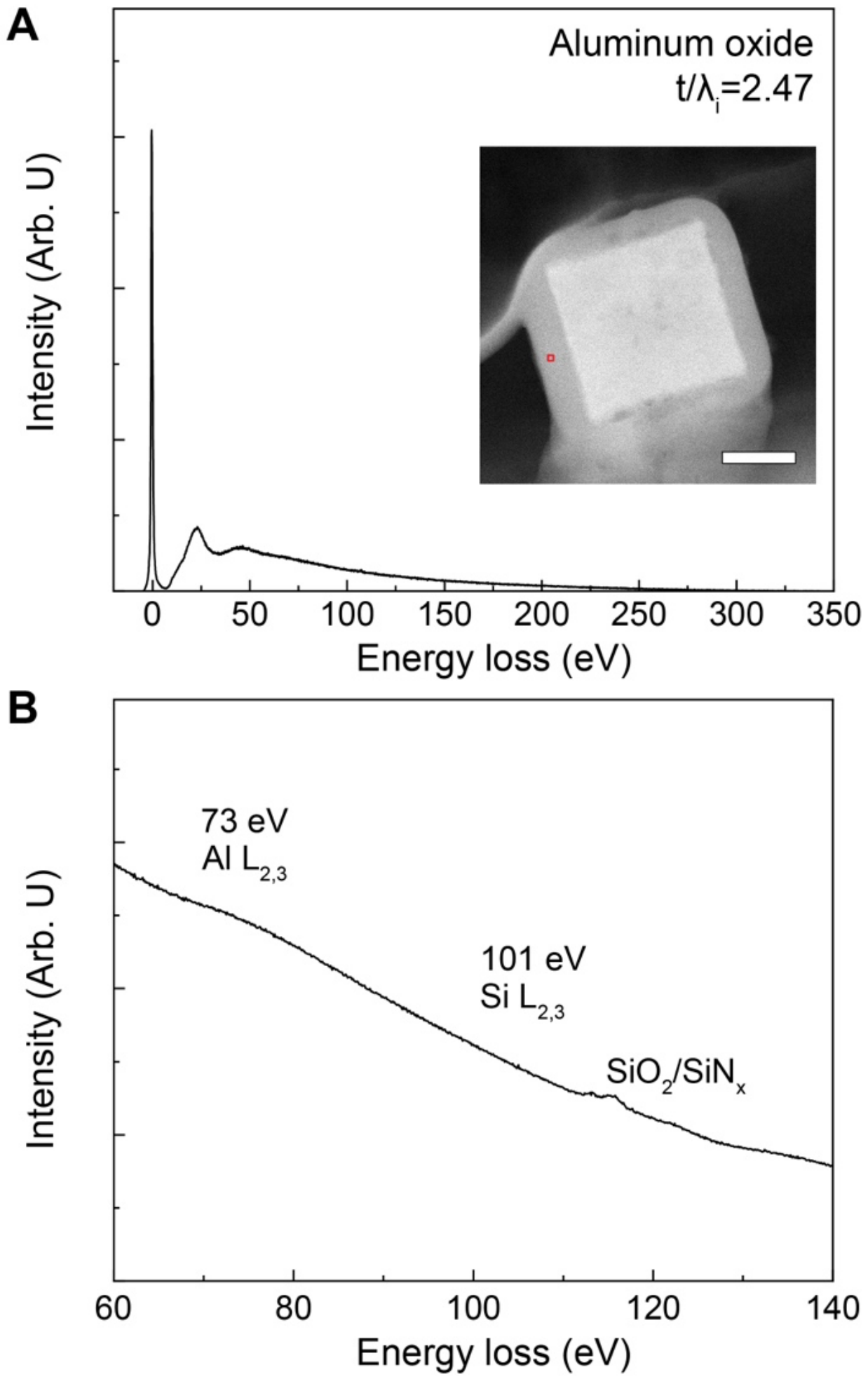

**Figure S12. EELS analysis of the ~200 nm aluminum nanocube oxide layer on a 30 nm $SiN_x$ membrane. (A)** Zero and low loss spectrum (0-350 eV). The red square dot in the inset TEM image indicates the location used for EELS analysis. The scale bar for inset image is 100 nm. **(B)** Core loss spectrum (60-140 eV). Negligible meaningful information can be retrieved from Al *L* core loss.

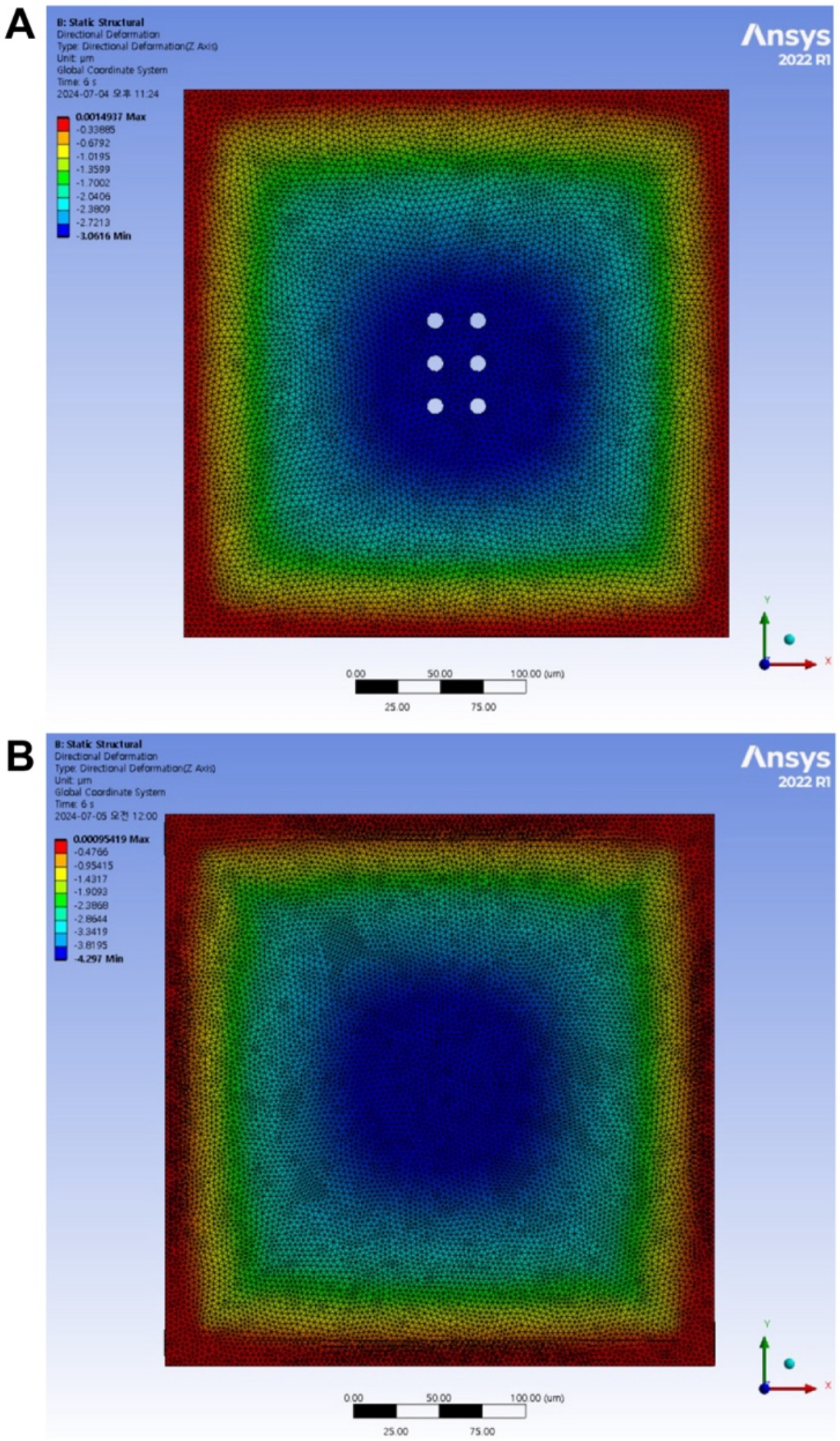

**Figure S13. Finite element analysis (FEA) of membrane deflection at 10 torr of pressure gradient.** (A) 150 nm-thick SiC heater membrane patterned with six 10 µm-diameter windows. The thickness of $SiN_x$ is 30 nm. (B) 50 nm $SiN_x$ window chip with a dimension of 300 × 300 µm. The detailed method for FEA is available in Supplementary Ref. 3

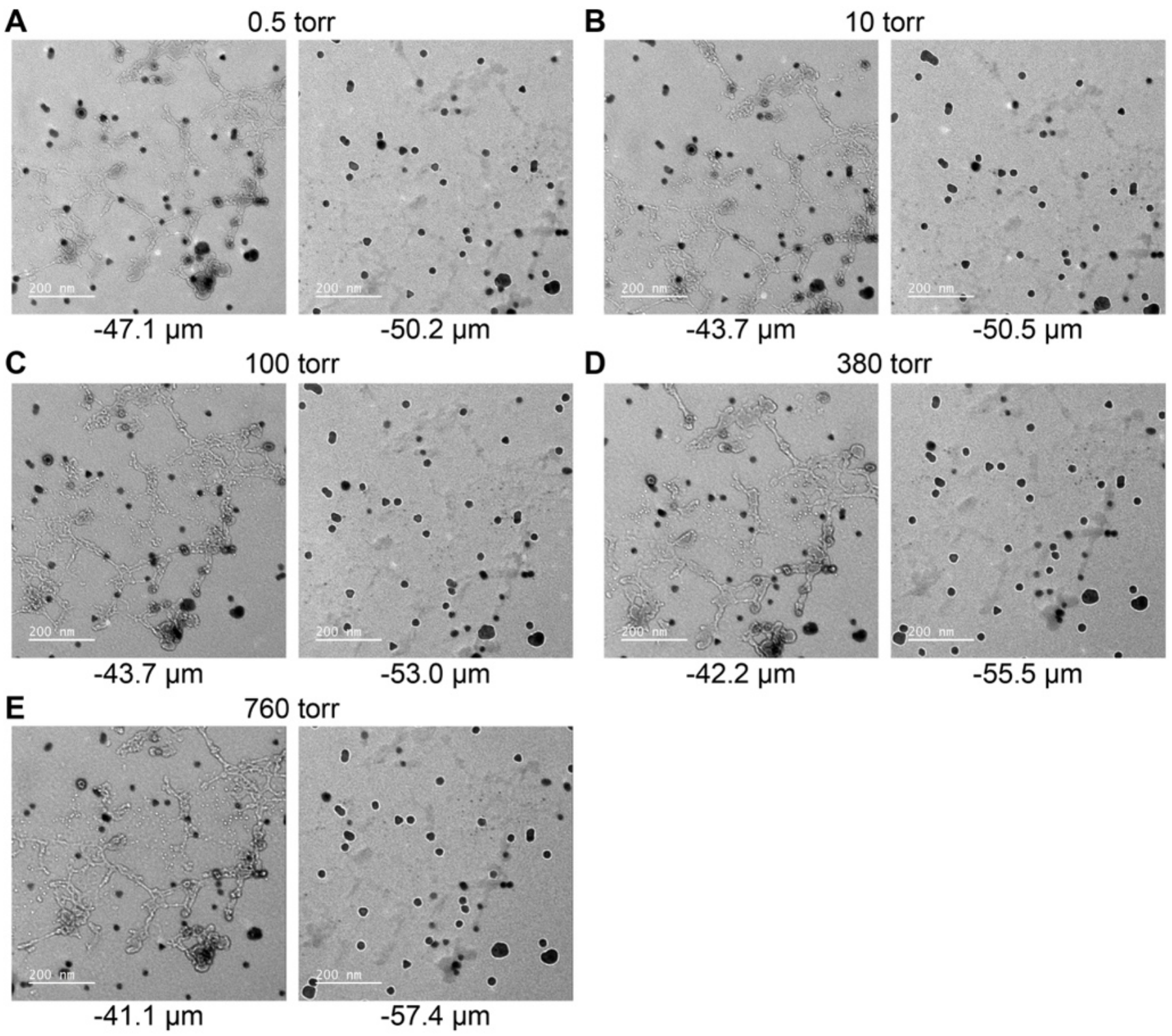

**Figure S14. Thickness measurement of the gas cell using the TEM defocus.** (A) at 0.5 torr, (B) at 10 torr, (C) at 100 torr, (D) at 380 torr, (E) at 760 torr Ar gas pressure.

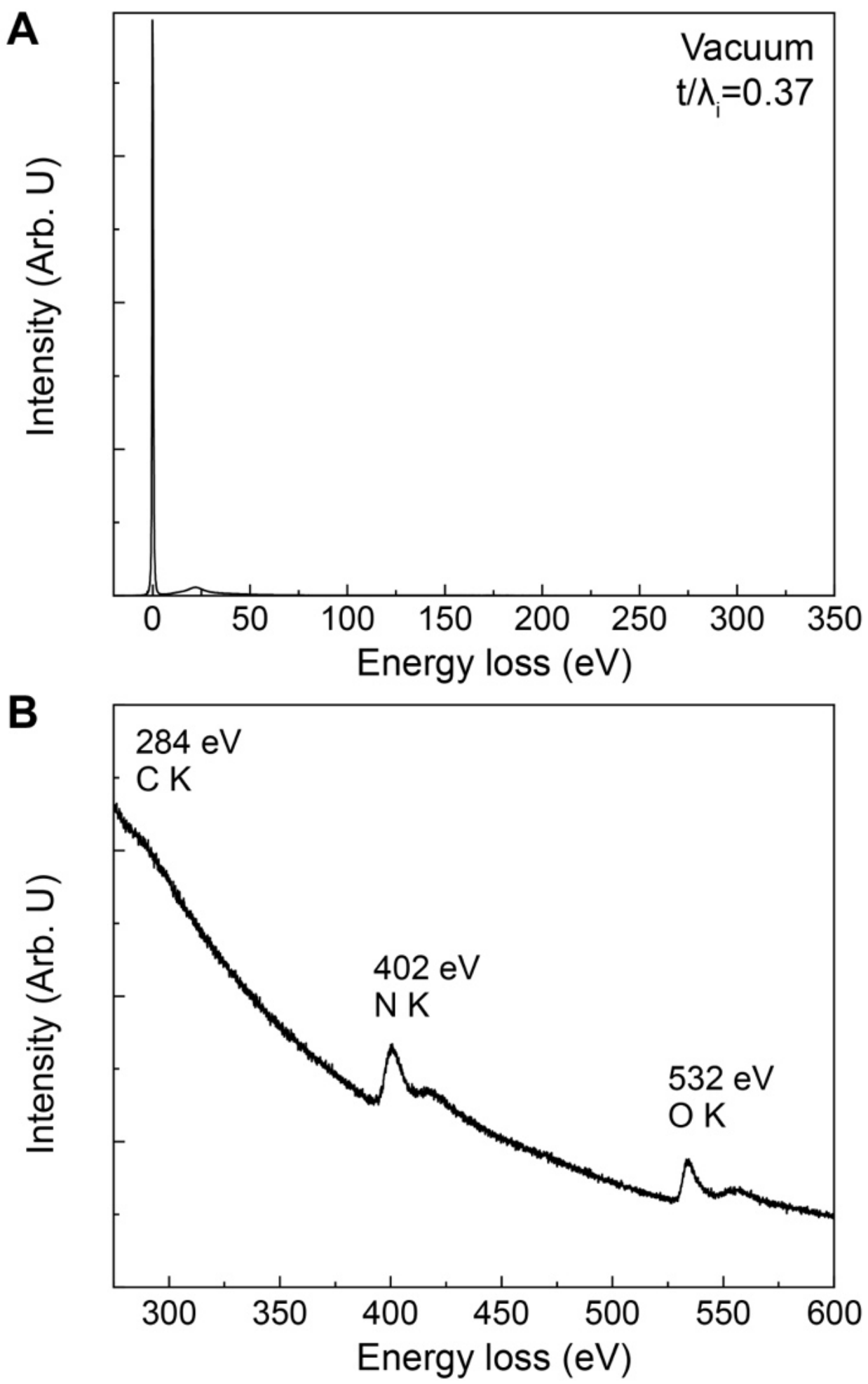

**Figure S15. EELS of the empty ultrathin silicon nitride gas cell. (A)** Zero and low loss spectrum (0-350 eV). **(B)** Core loss spectrum (275-600 eV). The core loss edge for nitrogen is from silicon nitride and that for oxygen is from partial oxidation of silicon nitride during the plasma cleaning. The electron dose rate is ~$1.2 \times 10^8$ Gy/s.

**Supplementary References**

1. Clark, B. D.; Jacobson, C. R.; Lou, M.; Renard, D.; Wu, G.; Bursi, L.; Ali, A. S.; Swearer, D. F.; Tsai, A.-L.; Nordlander, P.; Halas, N. J., Aluminum Nanocubes Have Sharp Corners. *ACS Nano* **2019,** *13* (8), 9682-9691.
2. Rosen, E. L.; Buonsanti, R.; Llordes, A.; Sawvel, A. M.; Milliron, D. J.; Helms, B. A., Exceptionally Mild Reactive Stripping of Native Ligands from Nanocrystal Surfaces by Using Meerwein's Salt. *Angewandte Chemie International Edition* **2012,** *51* (3), 684-689.
3. Koo, K.; Li, Z.; Liu, Y.; Ribet, S. M.; Fu, X.; Jia, Y.; Chen, X.; Shekhawat, G.; Smeets, P. J. M.; dos Reis, R.; Park, J.; Yuk, J. M.; Hu, X.; Dravid, V. P., Ultrathin Silicon Nitride Microchip for *in-situ*/*operando* Microscopy with High Spatial Resolution and Spectral Visibility. *Science Advances* **2024,** *10* (3), eadj6417.
4. Koo, K. kunmo-koo/Gas_radiolysis. https://github.com/kunmo-koo/Gas_radiolysis (accessed 2024/01/07).
5. Baird, J. K.; Miller, G. P.; Li, N., The G-value in Plasma and Radiation Chemistry. *Journal of applied physics* **1990,** *68* (7), 3661-3668.
6. Schneider, N. M.; Norton, M. M.; Mendel, B. J.; Grogan, J. M.; Ross, F. M.; Bau, H. H., Electron–water Interactions and Implications for Liquid Cell Electron Microscopy. *The Journal of Physical Chemistry C* **2014,** *118* (38), 22373-22382.
7. Hill, M.; Smith, F., Calculation of Initial and Primary Yields in the Radiolysis of Water. *Radiation physics and chemistry* **1994,** *43* (3), 265-280.
8. Willis, C.; Boyd, A.; Young, M.; Armstrong, D., Radiation Chemistry of Gaseous Oxygen: Experimental and Calculated Yields. *Canadian Journal of Chemistry* **1970,** *48* (10), 1505-1514.
9. Kummler, R.; Leffert, C.; Im, K.; Piccirelli, R.; Kevan, L.; Willis, C., A Numerical Model of Carbon Dioxide Radiolysis. *The Journal of Physical Chemistry* **1977,** *81* (25), 2451-2463.
10. Harteck, P.; Dondes, S., Radiation Chemistry of the Fixation of Nitrogen. *Science* **1964,** *146* (3640), 30-35.

11. Dondes, S.; Harteck, P.; Von Weyssenhoff, H., The Gamma Radiolysis of Carbon Monoxide in the Presence of Rare Gases. *Zeitschrift für Naturforschung A* **1964,** *19* (1), 13-18.

12. Berger, M. J.; Coursey, J. S.; Zucker, M. A.; Chang, J. NIST Stopping-Power and Range Tables: Electrons, Protons, Helium Ions http://www.nist.gov/pml/data/star/index.cfm/ (accessed Jun 28, 2024).

13. Fuller, E. N.; Schettler, P. D.; Giddings, J. C., New Method for Prediction of Binary Gas-phase Diffusion Coefficients. *Industrial & Engineering Chemistry* **1966,** *58* (5), 18-27.

14. Langenberg, S.; Carstens, T.; Hupperich, D.; Schweighoefer, S.; Schurath, U., Determination of Binary Gas-phase Diffusion Coefficients of Unstable and Adsorbing Atmospheric Trace Gases at Low Temperature–Arrested Flow and Twin Tube Method. *Atmospheric Chemistry and Physics* **2020,** *20* (6), 3669-3682.

15. Tang, M.; Cox, R.; Kalberer, M., Compilation and Evaluation of Gas Phase Diffusion Coefficients of Reactive Trace Gases in the Atmosphere: Volume 1. Inorganic Compounds. *Atmospheric Chemistry and Physics* **2014,** *14* (17), 9233-9247.

16. Ibraguimova, L.; Smekhov, G.; Shatalov, O. Recommended Rate Constants of Chemical Reactions in an $H_2$-$O_2$ Gas Mixture with Electronically Excited Species $O_2(\Delta)$, O(D), OH (2Σ) Involved. **2003**.

17. Willis, C.; Boyd, A.; Young, M., Radiolysis of Air and Nitrogen–Oxygen Mixtures with Intense Electron Pulses: Determination of a Mechanism by Comparison of Measured and Computed Yields. *Canadian Journal of Chemistry* **1970,** *48* (10), 1515-1525.

18. Solve 1-D Parabolic and Elliptic PDEs - MATLAB PDEPE. https://www.mathworks.com/help/matlab/ref/pdepe.html (accessed 2024/01/06).

19. Wang, M.; Park, C.; Woehl, T. J. Quantifying the Nucleation and Growth Kinetics of Electron Beam Nanochemistry with Liquid Cell Scanning Transmission Electron Microscopy. *Chemistry of Materials* **2018**, *30* (21), 7727-7736.